\documentclass{aa}  

\usepackage{graphicx}
\usepackage{txfonts}
\usepackage{lscape}
\begin{document}

   \title{Characterization and history of the Helmi streams with {\it Gaia} DR2}
   \author{Helmer H. Koppelman
         \inst{1}
         \and Amina Helmi \inst{1}
         \and Davide Massari \inst{1}
         \and Sebastian Roelenga \inst{1}
         \and Ulrich Bastian
          \inst{2}}

   \institute{Kapteyn Astronomical Institute, University of Groningen, Landleven 12, 9747 AD Groningen, The Netherlands\\
              \email{koppelman@kapteyn.astro.rug.nl}
              \and
              Zentrum für Astronomie, Heidelberg University, Astronomisches Rechen-Institut, Mönchhofstrasse 12-14, 69120 Heidelberg, \\ Germany}
         
   \date{Received -- --, ----; accepted -- --, ----}

  \abstract
  {The halo of the Milky Way has long been hypothesized to harbour
    significant amounts of merger debris. This view has been supported
    over more than a decade by wide-field photometric surveys which have
    revealed the outer halo to be lumpy.}
  {The recent release of {\it Gaia} DR2 is allowing us to establish
    that mergers also have been important and possibly built up the
    majority of the inner halo. In this work we focus on the Helmi
    streams, a group of streams crossing the Solar vicinity and known for
    almost two decades. We characterize their properties and
    relevance for the build-up of the Milky Way's halo.}
  {We identify new members of the Helmi streams in an unprecedented
    dataset with full phase-space information combining {\it Gaia}
    DR2, and the APOGEE DR2, RAVE DR5 and LAMOST DR4 spectroscopic
    surveys. Based on the orbital properties of the stars, we find new
    stream members up to a distance of 5 kpc from the Sun, which we characterize
    using photometry and metallicity information. We also perform
    N-body experiments to constrain the time of accretion and
    properties of the progenitor of the streams.}
  {We find nearly 600 new members of the Helmi streams. Their HR
    diagram reveals a broad age range, from approximately 11 to 13
    Gyr, while their metallicity distribution goes from $-2.3$ to
    $-1.0$, and peaks at [Fe/H]~$\sim -1.5$. These findings confirm
    that the streams originate in a dwarf galaxy. Furthermore, we find
    7 globular clusters to be likely associated, and which follow a
    well-defined age-metallicity sequence whose properties suggest a
    relatively massive progenitor object. Our N-body simulations
    favour a system with a stellar mass of
    $\sim 10^{8}\,\mathrm{M}_\odot$ accreted $5-8$ Gyr ago.}
  {The debris from the Helmi streams is an important donor to the
    Milky Way halo, contributing approximately 15\% of its mass in
    field stars and 10\% of its globular clusters. }


   \maketitle
%\[%\]
%________________________________________________________________

\section{Introduction}
According to the concordance cosmological model $\Lambda$CDM, galaxies grow by mass through mergers. Typically a galaxy's halo is formed through a handful of major mergers accompanied by a plethora of minor mergers. This model's predictions stem both from dark matter only simulations \citep{Helmi2003TheHalo} combined with semi-analytic models of galaxy formation \citep[e.g.][]{Bullock2005,Cooper2010} and hydrodynamical simulations \citep[e.g.][]{Pillepich2014}. 

When satellites merge with a galaxy like the Milky Way they get stripped of their stars by the tidal forces, e.g. \cite{Johnston1996FossilHalo}. These stars follow approximately the mean orbit of their progenitor and this leads to the formation of streams and shells. Wide-field photometric surveys have already discovered many cold streams likely due to globular clusters, e.g. Pal 5 \citep{Odenkirchen2001}; GD-1 \citep{Grillmair2006DETECTIONSURVEY}, and more disperse streams caused by dwarf galaxies, e.g. Sagittarius \citep{Ibata1994ASagittarius}, as well as large overdensities such as those visible in the SDSS or Pan-STARRS maps \citep{Belokurov2006,Bernard2016}. Many of these streams are distant and have become apparent after meticulous filtering \cite[e.g.][]{Rockosi2002,Grillmair2009}. 

Tidal debris in the vicinity of the Sun is predicted to be very
phase-mixed \citep{Helmi1999a,Helmi2003TheHalo}. Typically one can expect to find many stream wraps originating in the same object, i.e. groups of stars with different orbital phase sharing a common origin. Because of the high degree of phase-mixing, it is more productive to study tidal debris in spaces such as those defined by the velocities, integrals of motion \citep{Helmi2000MappingSatellites}, or in action space
\citep{McMillan2008DisassemblingCoordinates}, rather than to search for clustering in spatial coordinates. Until recently, a few studies on the nearby stellar halo identified different small groups of stars that likely were accreted together, e.g. \cite{Helmi1999,Chiba2000,Helmi2006,Klement2008,Klement2009a,Majewski2012EXPLORINGCENTAURI,Beers2017BRIGHTANALYSIS,Helmi2017}, see \cite{Smith2016TidalBeyond} for a comprehensive review.

A new era is dawning now that the {\it Gaia} mission is delivering full phase-space information for a billion stars. As it is clear from the above discussion, this is crucial to unravel the merger history of the Milky Way and to characterize the properties its progenitors. In fact, the second data release of the {\it Gaia} mission \citep{GaiaCollaboration2018brown} is already transforming the field of galactic archaeology.

The strength of {\it Gaia}, and especially of the 6D sample, is that it can identify stream members based on the measured kinematics \citep[e.g.][]{Koppelman2018OneDR2,Price-Whelan2018OffStream}. Arguably the most recent spectacular finding possible thanks to {\it Gaia} is the discovery that the inner halo was built largely via the accretion of a single object, as first hinted from the kinematics \citep{Belokurov2018Co-formationHalo,Koppelman2018OneDR2}, and the stellar populations \citep{GaiaCollaboration2018Babusiaux,Haywood2018InDR2}, all pieces put together in \cite{Helmi2018TheWay}. This accreted system known as Gaia-Enceladus was disky and similar in mass to the Small Magellanic Cloud today, and hence led to the heating of a proto-disk some 10 Gyr ago \citep{Helmi2018TheWay}.

Gaia-Enceladus debris however, is not the only substructure present in the vicinity of the Sun. Detected about 20 years ago, the Helmi streams \citep[H99 hereafter]{Helmi1999} are known to cross the Solar neighbourhood. Their existence has been confirmed by \cite{Chiba2000} and \cite{Smith2009}, among other studies. In the original work, 13 stars were detected based on their clumped angular momenta which clearly differ from other local halo stars. Follow-up work by \cite{Kepley2007HALONEIGHBORHOOD} estimated that the streams were part of the tidal debris of a dwarf galaxy that was accreted $6-9$ Gyr ago, based on the bimodality of the $z$-velocity distribution. This bimodal distribution is the distinctive feature of multiple wraps of tidal debris crossing the Solar neighbourhood. Since the discovery in 1999, a handful of new tentative members have been found increasing the total number of members to $\sim 30$ \citep[e.g.][]{Kepley2007HALONEIGHBORHOOD,ReFiorentin2005AstronomyStars,Klement2009a,Beers2017BRIGHTANALYSIS}, while several tens more were reported in \cite{GaiaCollaboration2018Helmi}. Also structure S2 from \cite{Myeong2017HaloClumps}, consisting of $\sim 60$ stars, has been recognized to be related to the Helmi streams \citetext{W.~Evans priv.\ comm.}.

Originally, the Helmi streams were found using Hipparcos proper motions \citep{Perryman1997} combined with ground-based radial velocities \citep{Beers1995KinematicsGalaxy,ChibaYoshii1998}. In this work, we aim to find new members and to characterize better its progenitor in terms of the time of accretion, initial mass and star formation history. To this end we focus on the dynamics, metallicity distribution and colour-magnitude diagram of its members. Furthermore, we also identify globular clusters that could have potentially been accreted with the object \citep{Leaman2013,DiederikKruijssen2018ThePopulation}, as for example seen for the Sagittarius dwarf galaxy \citep{Law2010ASSESSINGGALAXY,Massari2017The2419,Sohn2018AbsoluteMass}, and also for Gaia-Enceladus \citep{Myeong2018TheClusters,Helmi2018TheWay}.

This paper is structured as follows: in Section \ref{sec:data} we present the data and samples used, while in Section \ref{sec:moremembers} we define a core selection of streams members that serves as the basis to identify  more members. In Section \ref{sec:analysis} we analyze the spatial distribution of the debris. In Section \ref{sec:simulations} we supplement the observations with N-body simulations. We discuss possible associations of the Helmi streams with globular clusters in Section \ref{sec:globassoc}. Finally, we present our conclusions in Section \ref{sec:conclusion}.

\section{Data} \label{sec:data}
\subsection{Brief description of the data}
The recently published second data release (DR2) of the {\it Gaia}
space mission contains the on-sky positions, parallaxes, proper
motions, and the $G$, $G_\mathrm{BP}$ and $G_\mathrm{RP}$ optical
magnitudes for over 1.3 billion stellar
sources in the Milky Way \citep{GaiaCollaboration2018brown}. For
7,224,631 stars with $G_\mathrm{RVS}$ < 12, known as the 6D subsample,
line-of-sight velocity information measured by the {\it Gaia}
satellite is available \citep{Katz2018GaiaDR2}. The precision of the
observables in this dataset are unprecedented: the median proper
motions uncertainties of the stars with full phase-space information,
is 1.5 mas/yr which translates to a tangential velocity error of
$\sim 7$~km/s for a star at 1~kpc, while their median radial velocity
uncertainties are 3.3 km/s. This makes the {\it Gaia} DR2 both the
highest quality and the largest size single survey ever available for
studying the kinematics and dynamics of the nearby stellar halo and
disk.

\subsection{Cross-matching with APOGEE, RAVE and LAMOST}\label{sec:crossmatching}

To supplement the 6D {\it Gaia} subsample, we add the radial
velocities from the cross-matched catalogues APOGEE
\citep{Wilson2010TheSpectrograph,Abolfathi2018TheExperiment} and RAVE
DR5 \citep{Kunder2016}, see \cite{Marrese2018GaiaResults} for
details. We also add radial velocities from our own cross-match of
{\it Gaia} DR2 with LAMOST DR4 \citep{Cui2012}. 

For the cross-match with LAMOST we first transform the stars to the
same reference frame using the {\it Gaia} positions and proper
motions, and then we match stars within a radius of 10 arcsec with
TOPCAT/STILTS
\citep{Taylor2005TOPCATSoftware,Taylor2006STILTS-AData}. We find that
over 95\% of the stars have a matching radius smaller than 0.5 arcsec.
In total, we find 2,868,425 matches between {\it Gaia} and LAMOST DR4,
with a subset of 8,404 overlapping also with RAVE, and 50,650 with
APOGEE. Because the LAMOST radial velocities are known to be offset by
$+4.5$ km/s with respect to APOGEE \citep{Anguiano2018}, we correct
for this effect. 

Since the radial velocities of RAVE and APOGEE have been shown to be
very consistent with those of {\it Gaia}
\citep{Sartoretti2018GaiaData}, for our final catalogue, we use first
the radial velocities from APOGEE if available, then those from RAVE,
and finally from LAMOST for the stars for which there is no overlap
with either two of the other surveys. After imposing a quality cut of
{\tt parallax\_over\_error $> 5$}, this yields a sample of 2,361,519
stars with radial velocities. Note that all these surveys also provide
additional metallicity information for a subset of the stars.

When combined with the {\it Gaia} 6D sample, this results in a total
of 8,738,322 stars with 6D information and {\tt parallax\_over\_error
  $> 5$}. The median line-of-sight velocity error of the stars from
the ground-based spectroscopic surveys is 5.8 km/s, while that of the
pure {\it Gaia} sample is 1 km/s for the same parallax quality cut.

\subsection{Quality cuts and halo selection}
To isolate halo stars, we follow a kinematic selection, i.e. stars are
selected because of their very different velocity from local disk
stars. By cutting in velocity we introduce a clear bias: halo stars
with disk-like kinematics are excluded from this sample
\citep{Nissen2010,Bonaca2017GaiaHalo,Posti2018TheEllipsoid,Koppelman2018OneDR2}. Nevertheless,
the amplitude of the $Z$-velocities of the Helmi streams stars is
$> 200$ km/s, i.e. very different from the disk, so our selection
should not impact our ability to find more members.

\begin{figure*}
   \centering
   \includegraphics[width=\hsize]{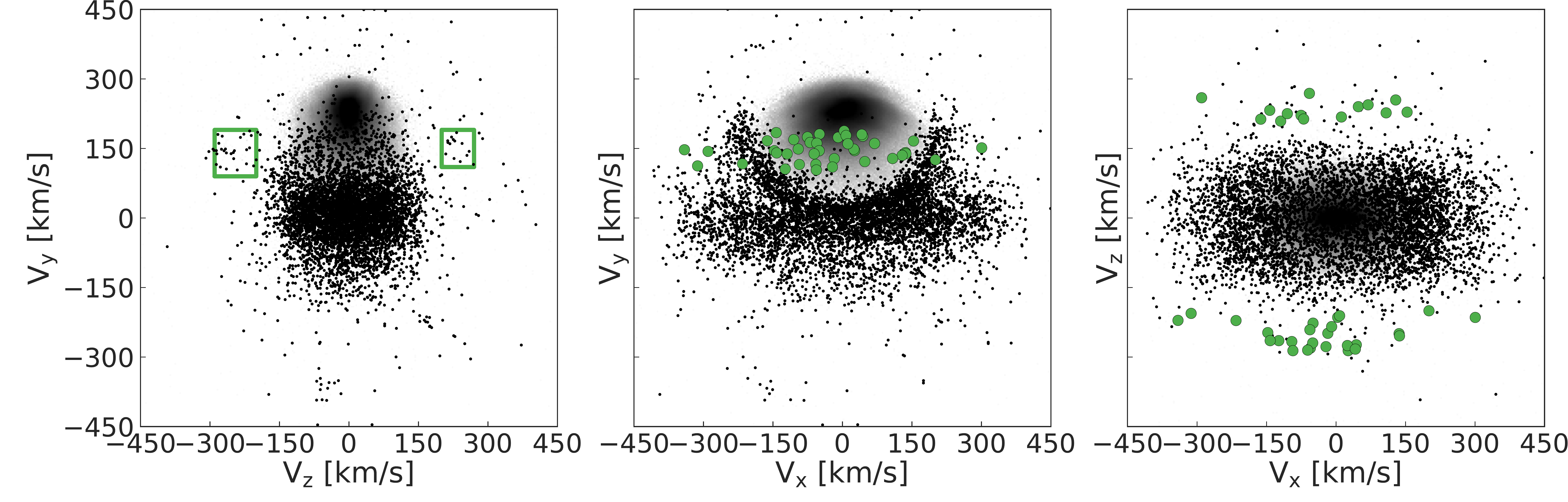}
   \caption{Velocity distribution of kinematically selected halo stars
     (black dots) within 1 kpc from the Sun from the {\it Gaia}-only
     6D sample. The grey density in the background shows the location
     and extent of the disk in this diagram. The velocities have been
     corrected for the motion of the Sun and LSR. The green boxes in
     the left panel indicate the location of the Helmi streams, and
     are drawn based on the velocities of the original stream
     members. The stars inside these boxes are highlighted with green
     symbols in the other two panels.}
         \label{fig:vxvy_kin_halo}
\end{figure*}
 
We start from our extended 6D data sample (obtained as described in
the previous section), and remove stars with {\tt
  parallax\_over\_error} > 5. Because of the zero-point offset of
$\sim -0.03$ mas known to affect the {\it Gaia} parallaxes
\citep{Arenou2018,GaiaCollaboration2018brown,Lindegren2018} we also
discard stars with parallaxes < 0.2~mas. Distances for this reduced sample
are obtained by inverting the parallaxes. Finally, following
\cite{Koppelman2018OneDR2} we select stars that have
$|\boldsymbol{V}-\boldsymbol{V_\mathrm{LSR}}| > 210$ km/s, where
$\boldsymbol{V_\mathrm{LSR}}$ is the velocity vector of the local
standard of rest (LSR). This cut is not too strict, allowing for some
contamination from the thick disk. This velocity selection is done
after correcting for the motion of the Sun using the values from
\cite{Schonrich2010LocalRest} and that of the LSR as estimated by
\cite{McMillan2017}. Our Cartesian reference frame is pointed such
that $X$ is positive toward the Galactic Centre, $Y$ points in the
direction of the motion of the disk, and $Z$ is positive for Galactic
latitude $b>0$. This final sample contains 79,318 tentative halo
stars, with 12,472 located within 1~kpc from the Sun, which is at
8.2~kpc from the Galactic Centre \citep{McMillan2017}. Slightly more
than half of these stars stem from the {\it Gaia}-only 6D sample.

\section{Finding members}\label{sec:moremembers}
\subsection{Core selection}
\label{sec:coremembers}

Using the halo sample described above, we will select `core members'
by considering only those stars within 1 kpc from the Sun from the {\it
  Gaia}-only sample. In such a local sample, streams are very
clustered in velocity-space because the gradients caused by the orbital
motion are minimized.

Fig.~\ref{fig:vxvy_kin_halo} shows with green boxes our selection of
the streams' core members in velocity space. The boxes are placed on
top of the positions of the original members of the Helmi streams. The
boundaries in $(V_Z,V_Y)$ for the left box are:
$[-290,-200]\,[90,190]$ km/s, and for the right box:
$[200,270]\,[110,190]$ km/s. 
Using the SIMBAD database we find that 10 of the original 13 members have {\it Gaia} DR2 distances smaller than $1$ kpc. Of these 10 stars, 9 have radial velocities in our extended data sample. Only one star with updated radial velocity information has
very different velocities, leaving the original sample with 8 reliable
members with full {\it Gaia} 6D parameters within $1$ kpc.

One of the key characteristics of the Helmi streams are the two groups
in the $V_Y-V_Z$ plane, one moving up through the disk and one moving
down. Using the selection described above we find 40 core members in
the {\it Gaia}-only sample, of which 26 with $V_Z<0$. The asymmetry in the
number of stars in the two streams can be used as an indicator of the
time of accretion and/or mass of the progenitor since tidal streams
will only produce multiple wraps locally after having evolved for a
sufficiently long time. In \S \ref{sec:progprop} we explore what the
observed asymmetry implies for the properties of the progenitor of the
Helmi streams.

\begin{table}
  \caption{{\it Gaia} DR2 source\_id of the Helmi streams' core members.}             % title of Table
\label{table:1}      % is used to refer this table in the text
\centering                          % used for centering table
\begin{tabular}{c c}        % centered columns (4 columns)
\hline\hline                 % inserts double horizontal lines
IDs members 1-20 & IDs members 21-40 \\    % table heading 
\hline                        % inserts single horizontal line
365903386527108864 & 604095572614068352 \\
640225833940128256 & 1049376272667191552 \\
1415635209471360256 & 1621470761217916800 \\
1639946061258413312 & 2075971480449027840 \\
2081319509311902336 & 2268048503896398720 \\
2322233192826733184 & 2416023871138662784 \\
2447968154259005952 & 2556488440091507584 \\
2604228169817599104 & 2670534149811033088 \\
2685833132557398656 & 2891152566675457280 \\
3085891537839264896 & 3085891537839267328 \\
3202308378739431936 & 3214420461393486208 \\
3306026508883214080 & 3742101345970116224 \\
4440446153372208640 & 4768015406298936960 \\
4998741805354135552 & 5032050552340352384 \\
5049085217270417152 & 5388612346343578112 \\
5558256888748624256 & 5986385619765488384 \\
6050982889930146304 & 6170808423037019904 \\
6221846137890957312 & 6322846447087671680 \\
6336613092877645440 & 6545771884159036928 \\
6615661065172699776 & 6914409197757803008 \\
\hline                                   %inserts single line
\end{tabular}
\end{table}

\subsection{Beyond the core selection}\label{sec:beyondcore}
The original 13 H99 members are located within 2.5 kpc from the
Sun. Beyond this distance, we expect that other members of the streams
will have different kinematic properties because of velocity gradients
along their orbit. Therefore the best way of finding more members
beyond the local volume is to use integrals of motion (IOM) such as
angular momenta and energy or the action integrals
\citep{Helmi2000MappingSatellites}. Note that from this section
onwards, we use the extended sample described in
Section \ref{sec:crossmatching} and which includes radial velocities from
APOGEE/RAVE/LAMOST.

Here we will mainly base the membership selection on the angular
momentum of the stars, namely the $z$-component $L_z$, and the
perpendicular component: $L_\perp = \sqrt{L_x^2+L_y^2}$, although the
latter is generally not fully conserved. Since the energy $E$ depends
on the assumed model for the galactic potential, we use it only to
check for outliers.

To calculate the energy of the stars we model the Milky Way with a
potential that is similar to that used by \cite{Helmi2017}: it
includes a Miyamoto-Nagai disk with parameters
$\mathrm{M}_d = 9.3\cdot 10^{10}\,\mathrm{M}_\odot$,
$(a_d,b_d) = (6.5,0.26)\,\mathrm{kpc}$, an NFW halo with parameters
$\mathrm{M}_h = 10^{12}\,\mathrm{M}_\odot$,
$r_{s,h} = 21.5\,\mathrm{kpc}$, $c_h = 12$, and a Hernquist bulge with
parameters $\mathrm{M}_b = 3\cdot10^{10}\,\mathrm{M}_\odot$,
$c_b = 0.7$ kpc. The circular velocity curve of this potential is
similar to that of the Milky Way. Since this potential is axisymmetric
$L_z$ is a true IOM. Note that for convenience, in what follows we
flip the sign of $L_z$ such that it is positive for the Sun.

\begin{figure}
   \centering
   \includegraphics[width=\hsize]{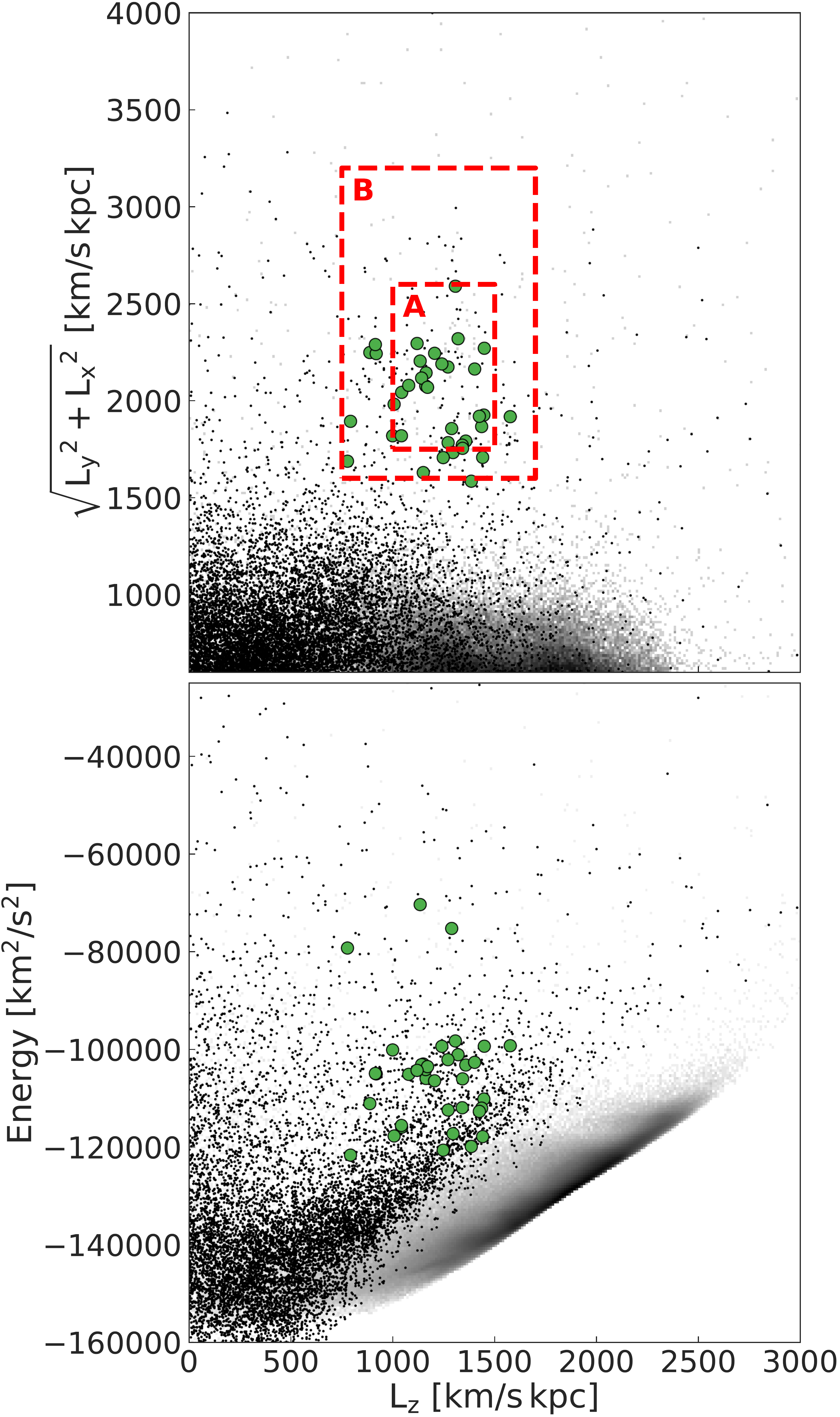}
   \caption{Distribution of the core members of the Helmi streams
     (green symbols), in angular momentum $L_\perp$ and $L_z$ in the
     top panel, and energy $E$ and $L_z$ in the bottom panel. The
     black dots show all of the stars of our sample of kinematically selected
     halo stars, located within 2.5 kpc from the Sun. With red, dashed
     lines we indicate the limits we apply to select additional
     members of the Helmi streams. In the background, the location of
     the disk(s) is shown with a grey density map.}
   \label{fig:IOMselection}
\end{figure}

In Fig.~\ref{fig:IOMselection} we show the distribution of the core
members in $L_z$ versus $L_\perp$ (top), and $L_z$ versus $E$
(bottom). A grey density map of all the stars in the {\it Gaia} 6D
sample with 20\% relative parallax error and with parallax $> 0.2$
illustrates the location of the disk. The black dots correspond to all
kinematically selected halo stars within 2.5 kpc from the Sun. The
transition of the halo into the disk is smooth, and only appears sharp
because of this particular visualization. The green dots are the core
members selected in \S~\ref{sec:coremembers}. These are clumped
around $(L_\perp,L_z) \sim (2000,1250)$~kpc~km/s. With red, dashed
lines we indicate two boxes labelled A \& B, that we use to select
tentative additional stream members. The limits of box A are:
$1750<L_\perp<2600$~kpc~km/s and $1000<L_z<1500$~kpc~km/s, and those
of box B are: $1600<L_\perp<3200$~kpc~km/s and
$750<L_z<1700$~kpc~km/s. Box B allows for more members, but also for
more contamination from the local halo background or thick disk. In
the bottom panel, we see that some of the core members appear to be
outliers with too high or low energy (but the analyses carried out in
the next sections show they are indistinguishable in their other
properties, except for their large $v_R$ velocities, see
\S~\ref{sec:progprop}). 

The number of stars located within 5 kpc that fall in boxes A \& B are
respectively 235 and 523. Note that for these selections we have used
the full extended sample without a kinematical halo selection. At most
20 stars that we identify as members of the Helmi streams in the IOM
space (grey points inside the selection boxes), do not satisfy our
halo selection (meaning that they have
$|\boldsymbol{V}-\boldsymbol{V_\mathrm{LSR}}| < 210$ km/s).

In the following sections, we focus on the stream members that fall in
selection B unless mentioned otherwise. We remind the reader that
selection B includes stars from the full extended sample comprising
radial velocities from {\it Gaia} and from APOGEE/RAVE/LAMOST. A table
listing all the members in selection B can be found in the Appendix.

\subsection{Members without radial velocities}\label{sec:5Dmembers}

Most of the stars in the {\it Gaia} DR2 dataset lack radial
velocities and this makes the search for additional tentative members
of the Helmi streams less straightforward. In
\cite{GaiaCollaboration2018Helmi} new members were identified using
locations on the sky where the radial velocity does not enter in the
equations for the angular momentum, namely towards the Galactic centre
and anti-centre. At those locations, the radial velocity is aligned
with the cylindrical $v_R$ component, therefore, it does not
contribute to $L_z = r v_\phi$, and $L_y = -xv_z$. The degree to which
the radial velocity contributes to the angular momenta increases with
angular distance from these two locations on the sky. Based on a
simulation of a halo formed through mergers created by
\cite{Helmi2000MappingSatellites}, we estimate that within 15 degrees
from the (anti) centre, the maximum difference between the true
angular momenta of stars and that computed assuming a zero radial
velocity, is $\sim$ 1000~kpc~km/s. Since the size of Box B is
$\sim 1000$~kpc km/s, we consider 15 degrees as the maximum tolerable
search radius. We denote the angular momenta computed assuming zero
radial velocity as $\widetilde{L}_y$ and $\widetilde{L}_z$, where we
change the sign of $\widetilde{L}_z$ such that it is positive in the
(prograde) direction of rotation of the disk.

\begin{figure}
   \centering
   \includegraphics[width=\hsize]{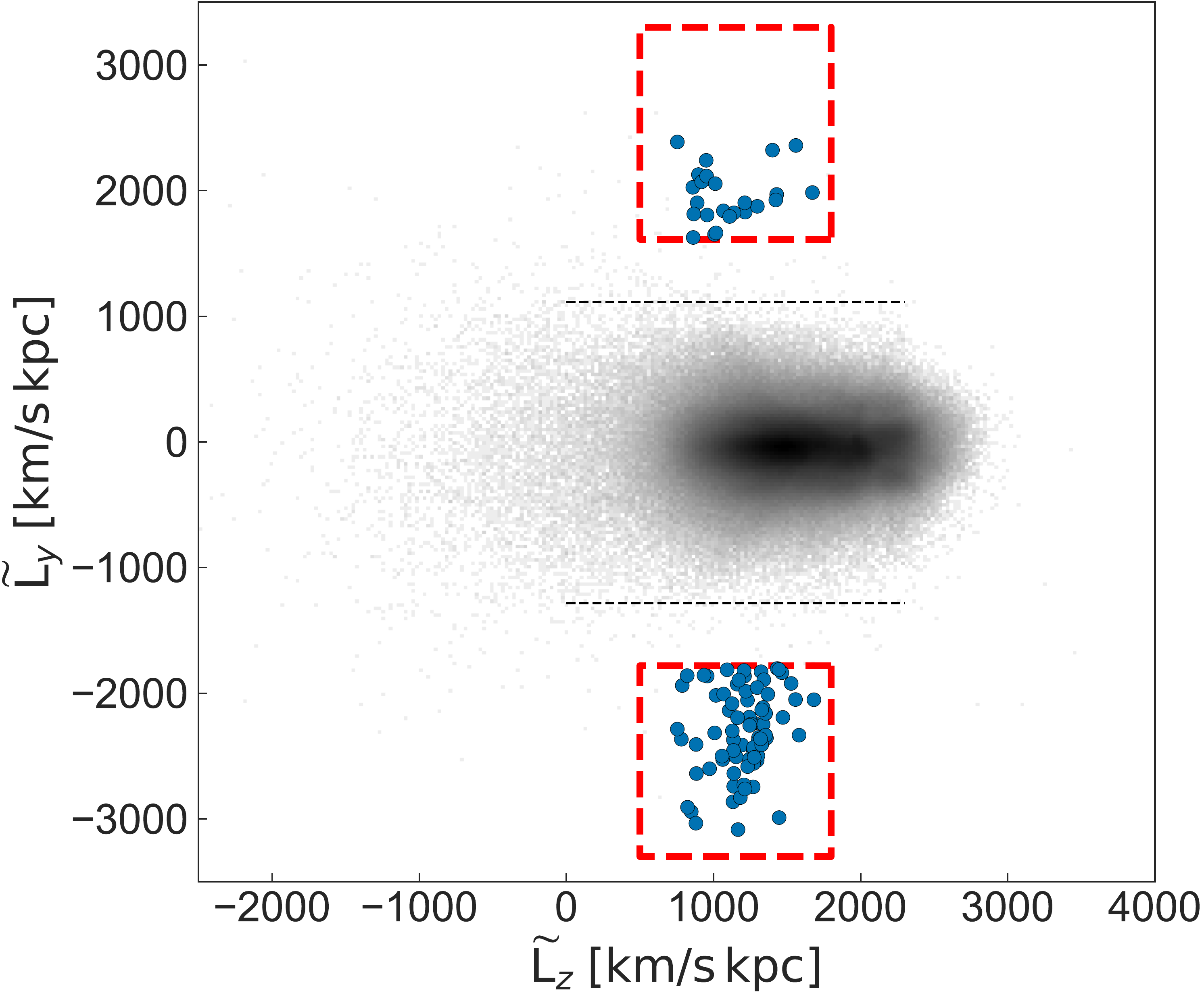}
   \caption{Distribution of stars from the 5D subset of {\it Gaia},
     located within 15 degrees from the Galactic centre or
     anti-centre, in (pseudo)angular momentum space. The angular
     momenta are calculated here by assuming that the line-of-sight
     velocities are zero. The grey density map reveals the location of
     the stars, most of which are in the disk. The black dashed lines
     show the 2.5\% and 97.5\% quantiles of the
     $\widetilde{L}_y$-distribution. The two boxes indicated with red
     dashed lines are used to identify candidate members of the Helmi
     streams, here shown with blue symbols.}
   \label{fig:5Dselection}
\end{figure}

Therefore, using the full {\it Gaia} DR2 5D-dataset, we select stars
within 15 degrees from the (anti)centre and with {\tt
  parallax\_over\_error > 5}, and apply the following photometric
quality cuts described in \S2.1 of
\cite{GaiaCollaboration2018Babusiaux}: {\tt
  phot\_g\_mean\_flux\_over\_error} > 50, {\tt
  phot\_rp\_mean\_flux\_over\_error} > 20, \linebreak {\tt
  phot\_bp\_mean\_flux\_over\_error} > 20, 1.0 + 0.015$\times$ power({\tt
  phot\_bp\_mean\_mag-phot\_rp\_mean\_mag},2) < {\tt
  phot\_bp\_rp\_excess\_factor} < 1.3 + 0.06$\times$ power({\tt
  phot\_bp\_mean\_mag-phot\_rp\_mean\_mag},2).
Fig.~\ref{fig:5Dselection} shows the distribution of all the stars in
this subsample with a grey density map in $\widetilde{L}_y$ vs
$\widetilde{L}_z$ space. The two boxes marked with red dashed lines
show the criteria we apply to identify additional members of the Helmi
streams. The size and location of the boxes are based on those in
Fig.~\ref{fig:IOMselection}. They are limited by
$750<\widetilde{L}_z<1700$~kpc~km/s, while we use a tighter constraint
on $|\widetilde{L}_y|$ to prevent contamination from the disk. The
black dashed lines in Fig.~\ref{fig:5Dselection} indicate upper and
lower quantiles of the full $\widetilde{L}_y$-distribution such that
95\% of the stars in the 5D subsample are located between these dashed
lines. The lower limits of selection boxes in the $\widetilde{L}_y$
direction are offset by 500~kpc~km/s from the dashed lines, and are
located at $\widetilde{L}_y$ at 1782 and -1613~kpc~km/s, respectively.

The blue symbols in Fig.~\ref{fig:5Dselection} correspond to the 105
tentative members that fall inside the boxes. Most of these stars are
within 2.5 kpc from the Sun. The two clumps in $\widetilde{L}_y$
have a direct correspondence to the two streams seen in the $V_Z$
component for the 6D sample. The clumps have 24 and 81 stars each,
implying a $\sim$ 1:3 asymmetry which is quite different from that
seen in the number of core member stars associated with each of the two
velocity streams. The difference could be caused in part by
incompleteness and crowding effects together with an anisotropic
distribution of the stars in the streams (see e.g.
Fig.~\ref{fig:orbits}). We use the 5D members in this work only for
the photometric analysis of the Helmi streams carried out in
Section \ref{sec:phot}.

%###################################################################
%-------------------------------------------------------------------
%###################################################################

\section{Analysis of the streams}\label{sec:analysis}
\subsection{Spatial distribution}
\begin{figure}%[!ht]
   \centering
   \includegraphics[width=0.85\hsize]{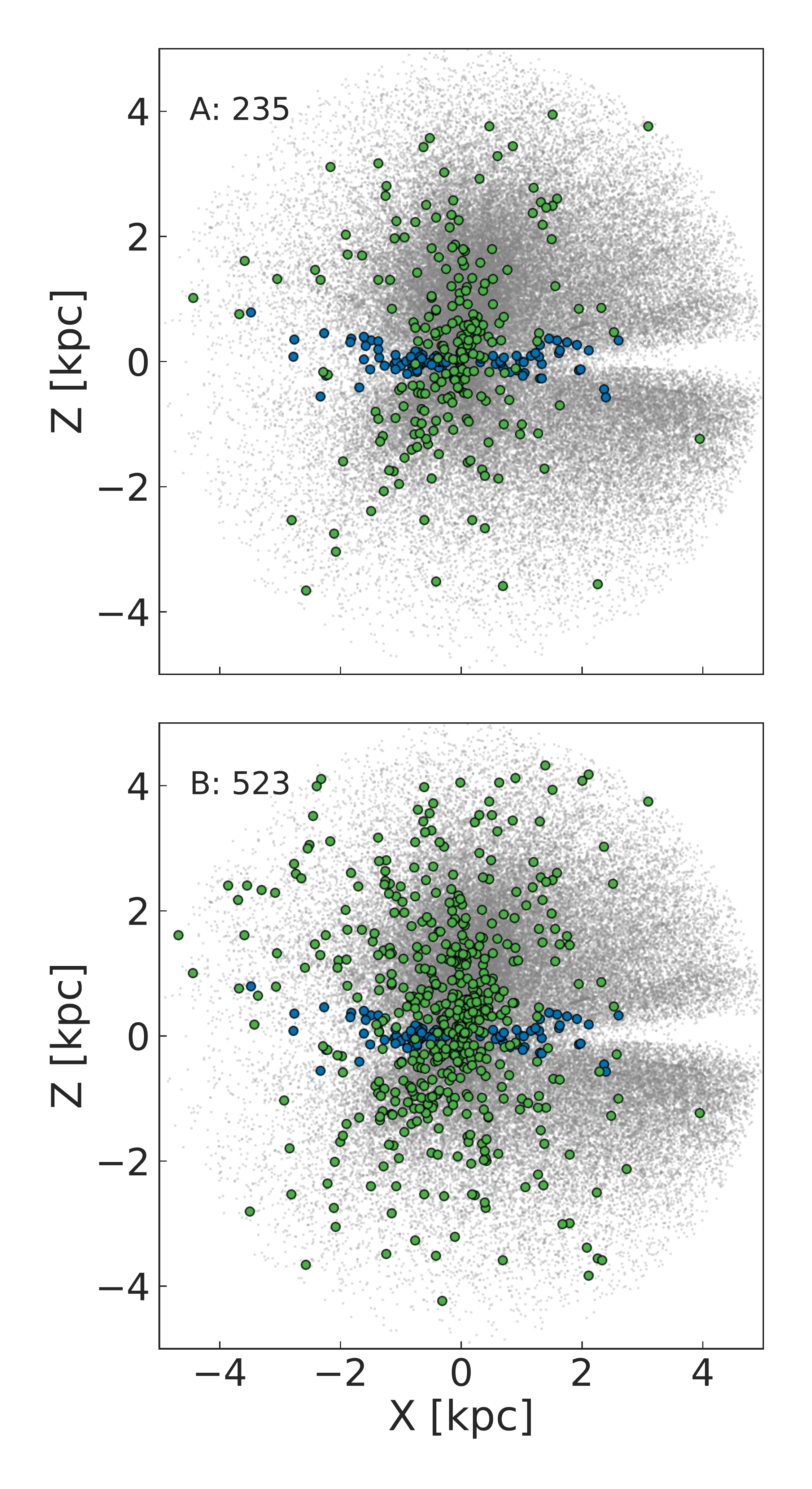}
   \caption{Distribution of members of the Helmi streams for the two
     selection boxes A (top) and B (bottom). The 6D stream members are
     indicated with green circles, local halo stars are shown in the
     background with grey symbols. The total number of 6D stream
     members is indicated in the top left of each
     panel. Tentative members from the 5D data set are shown with blue
     symbols.}
   \label{fig:xz_various}
\end{figure}

Fig.~\ref{fig:xz_various} shows the distribution of the streams
members in the $XZ$-plane, for the selection box A (top) and for B
(bottom). Those identified with 6D information are indicated with
green circles, a local sample of halo stars is shown in grey in the
background. There is a lack of stars close to the plane of the disk,
likely due to extinction \citep{Katz2018GaiaDR2}. This gap is filled
with tentative members from the 5D sample (in blue) which have, by
construction, low galactic latitude. Fig.~\ref{fig:xz_various} reveals
the streams stars to be extended along the $Z$-axis, as perhaps
expected from their high $V_Z$ velocities, but there is also a clear
decrease in the number of members with distance from the Sun.
\begin{figure}%[!ht]
   \centering
   \includegraphics[width=\hsize]{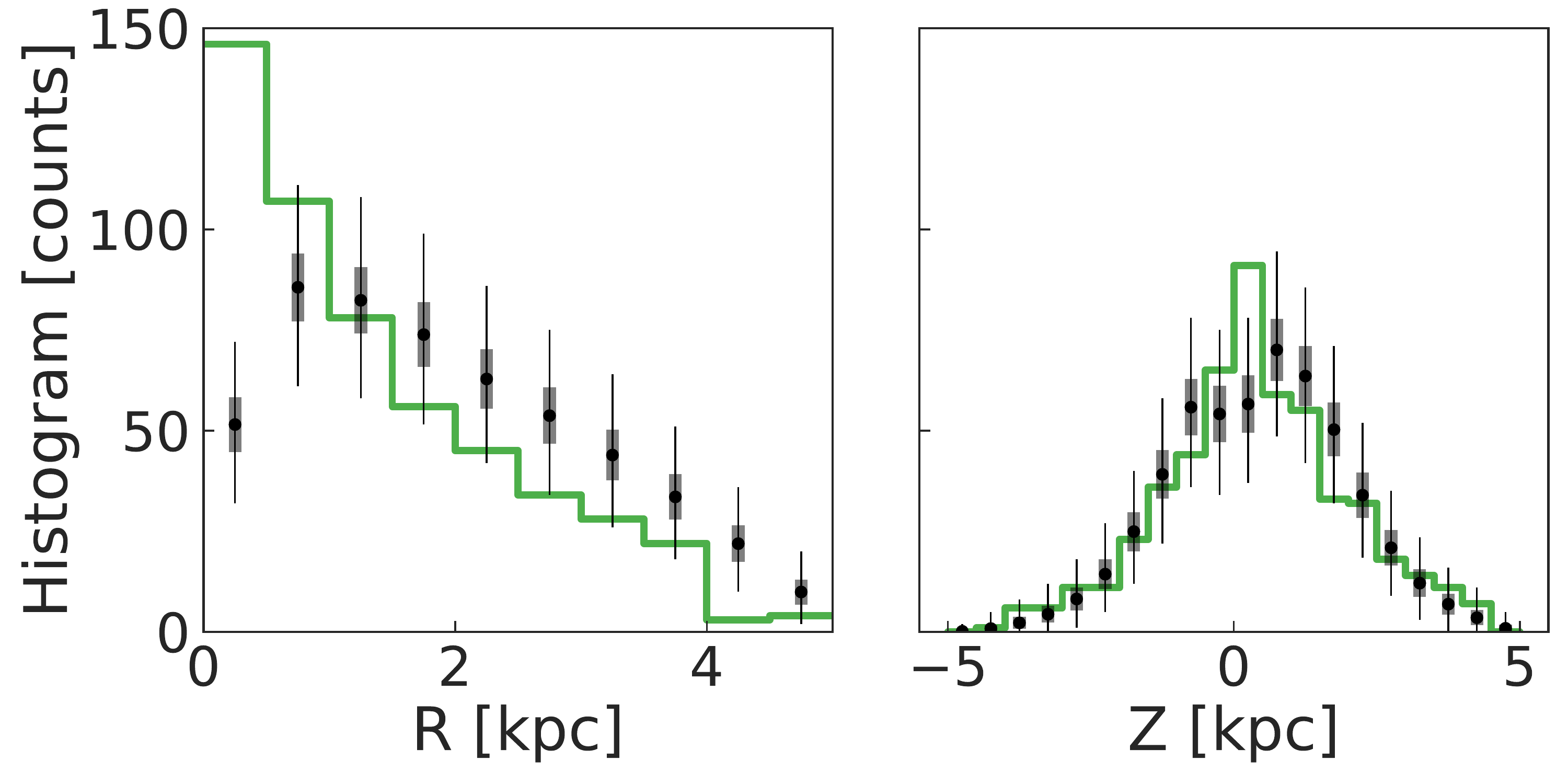}
   \caption{Distribution of heliocentric distance in the plane of the
     disk (left) and height above the plane (right) for stars in our
     6D sample. The green histograms are for members of the Helmi
     streams.  With black symbols, we show the mean counts obtained
     using $10^4$ random sets extracted from our (background) halo
     sample, with the grey and black error bars indicating the
     1$\sigma$ and 3$\sigma$ uncertainties for each $R/Z$-bin. The Helmi
     streams are clearly more confined in the $R$-direction.}
   \label{fig:RZ_random}
\end{figure}

\begin{figure*}
\centering
  \includegraphics[width=16cm]{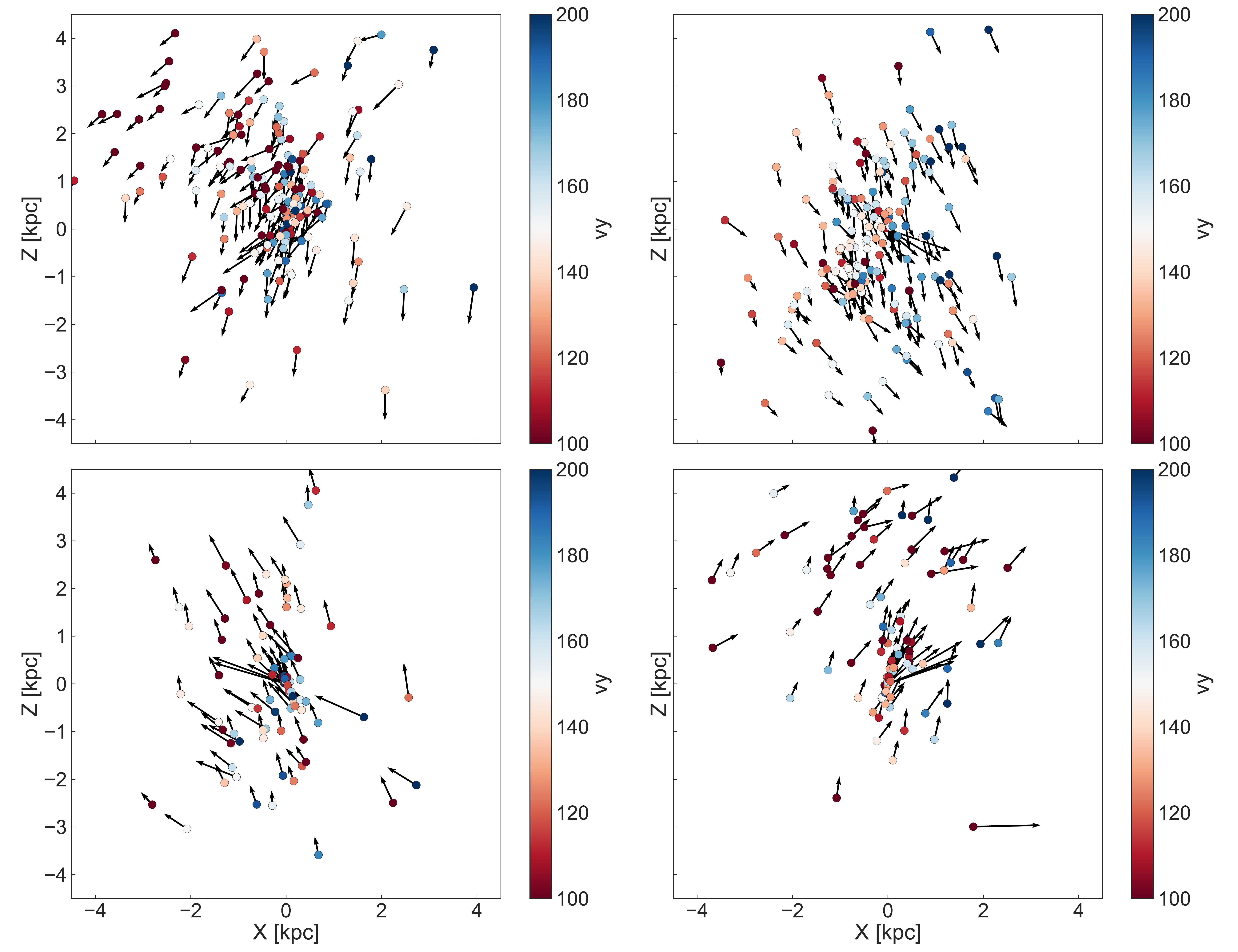}
  \caption{Distribution of the stars in the Helmi streams in the $XZ$
    plane (i.e. perpendicular to the disk of the Milky Way), with the
    arrows illustrating their motions (amplitude and direction) in
    this plane. The symbols are colour-coded according to the
    amplitude of their $V_Y$ velocity (i.e. perpendicular to the
    projected plane). The top and bottom rows show stream members with
    $V_Z<0$ and $V_Z>0$, respectively. The left and right columns show
    stars with $V_X<0$ and $V_X>0$, respectively. In all the four
    panels streaming motions and substructures are clearly apparent.}
  \label{fig:spatial_distribution_XZ}
\end{figure*}

To establish whether the spatial distribution of the streams differs
from that of the background, we proceed as follows. We compare our
sample of streams stars to $10^4$ samples randomly drawn from the
background. The random samples contain the same number of stars as the
streams, and the background comprises all of the stars in the halo sample,
described in Section \ref{sec:data}, excluding the streams members. In this
way, we account for selection effects associated with the different
footprints of the APOGEE/RAVE/LAMOST surveys as well as with the 20\%
relative parallax error cut (since the astrometric quality of the {\it
  Gaia} data is not uniform across the sky), and which are likely the
same for the streams and the background.

Figure \ref{fig:RZ_random} shows a comparison of the distribution of
heliocentric $R$ (left) and $Z$ (right) coordinates of the stars in
our sample (in green) and in the random samples (black). The sizes of the
grey and black markers indicate the 1$\sigma$ and 3$\sigma$ levels
respectively, of the random samples. The $Z$-distribution of the
streams members shows minimal differences with respect to the
background, except near the plane, i.e. for $Z \sim 0$.  On the other hand,
their distribution in $R$ shows very significant differences with
respect to the background, depicting a very steeply declining
distribution. This was already hinted at in
Fig.~\ref{fig:spatial_distribution_XZ}, and would suggest that the
streams near the Sun have a cross section\footnote{defined as the
  distance at which the counts of stars has dropped by a factor two.}
of $\sim 500$~pc.

\subsection{Flows: velocity and spatial structure}

The main characteristic of stars in streams is that they move together
through space as in a flow. Figure \ref{fig:spatial_distribution_XZ}
illustrates this by showing the spatial distribution of the members
(according to selection B), in the $XZ\mbox{-}$plane. The arrows
indicate the direction and amplitude of the velocities of the stars,
with stars with $V_Z < 0$ shown in the top row and those with
$V_Z > 0$ in the bottom row of the figure. Every star is colour-coded
according to its velocity component in and out of the plane of
projection (i.e. its $V_Y$). The left column shows stars with
$V_X < 0$, while the right column shows stars with $V_X > 0$.

The flows seen in Fig.~\ref{fig:spatial_distribution_XZ} reveal that
the two characteristic clumps in $V_Z$ (i.e. those shown in the left
panel of Fig.~\ref{fig:vxvy_kin_halo}) actually consist of several
smaller streams. For example, the top and bottom right panels of
Fig.~\ref{fig:spatial_distribution_XZ} both clearly show two flows:
one with $V_X \sim 0$, the other with a large $V_X$.

\begin{figure}
   \centering
   \includegraphics[width=\hsize]{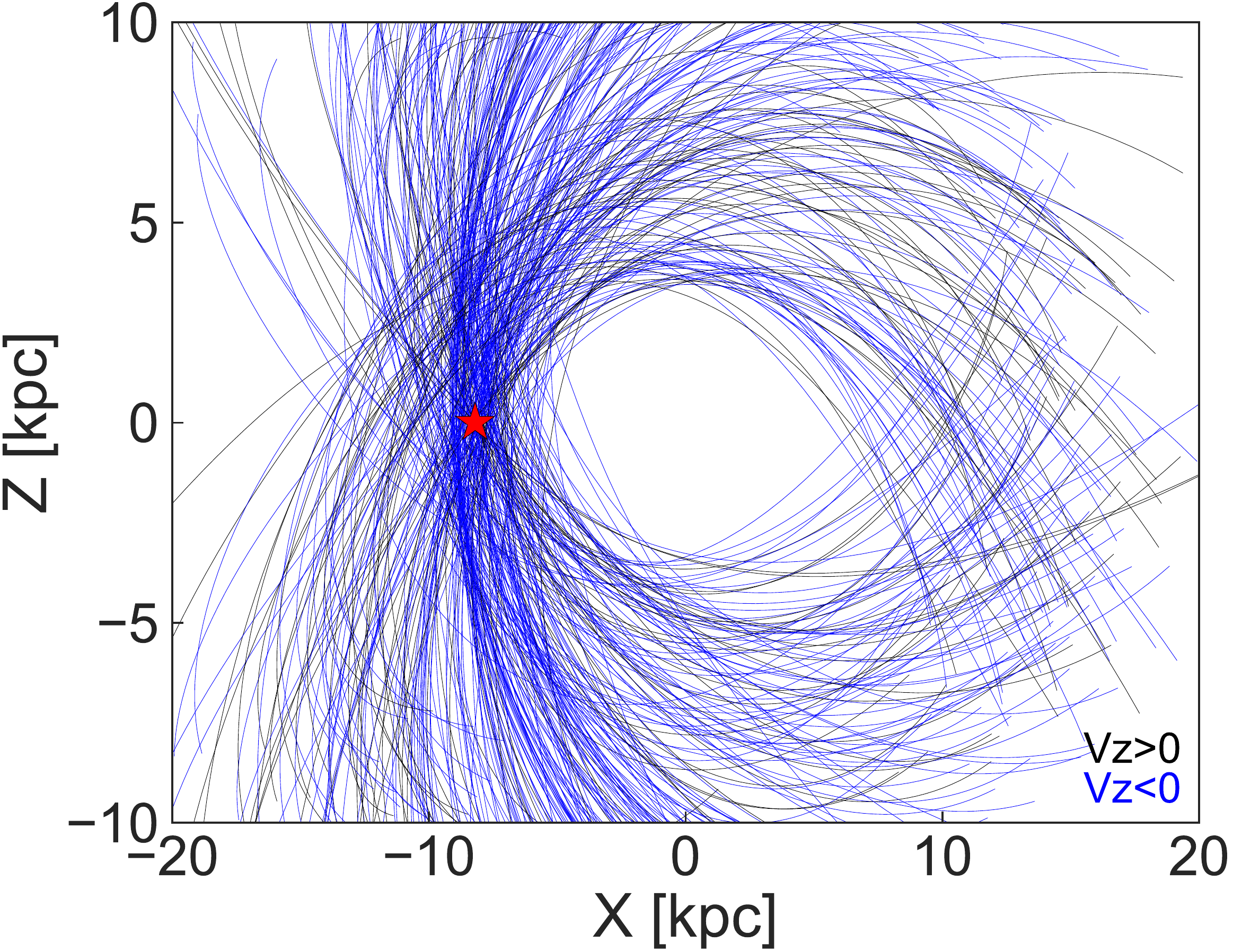}
   \caption{Compilation of orbits based on the 6D positions of the
     members of the Helmi streams. These orbits have been integrated
     for 100 Myr backwards and forward in time. The position of the
     Sun is illustrated with a red star, the Galactic centre is at
     $(X,Z) = (0,0)$ in this frame. The trajectories of stars that
     currently have $V_Z>0$ are coloured black, while those with
     $V_Z<0$ are shown in blue. Close to the Solar position, the
     majority of the Helmi streams' members move perpendicular to the
     plane of the disk, and are close to pericentre.}
         \label{fig:orbits}
\end{figure}

To enhance the visibility of the flows we integrate the orbits of the
streams stars forward and backwards in time. The potential in which
the orbits are calculated is the same as the one described in
Section \ref{sec:beyondcore}. The trajectories of all the stars are
integrated for $\pm 100$ Myr in time and are shown in
Fig.~\ref{fig:orbits} projected onto the $XZ$ plane. With a red star,
we indicate the Solar position. The trajectories of stars that belong
to the group with $V_Z > 0$ are coloured black, while
those with $V_Z < 0$ are given in blue.

Clearly, the stars found in the Solar vicinity are close to an orbital
turning point and on trajectories elongated in the $Z$-direction, as
expected from their large vertical
velocities. Fig.~\ref{fig:orbits} serves also to understand the
observed spatial distribution of the member stars (i.e. narrower in
$X$ (or $R$) and elongated in $Z$) seen in
Fig.~\ref{fig:xz_various}. Finally, we also note the presence of groups
of orbits tracing the different flows just discussed, such as for
example the group of stars moving towards the upper left corner of the
figure (and which corresponds to some of the stars shown in the bottom
left panel of Fig.~\ref{fig:spatial_distribution_XZ}).

\subsection{Ratio of the number of stars in the two $V_Z$
  streams}\label{sec:ratiodata}

As mentioned in the introduction, the ratio of the number of stars in
the two $V_Z$ streams was used in \cite{Kepley2007HALONEIGHBORHOOD} to
estimate the time of accretion of the object.  The ratio these authors
used was 1:2, in good agreement with the ratio found here in \S
\ref{sec:coremembers} for the core members. Using numerical
simulations, this implied an accretion time of $6-9$ Gyr for an object
of total dynamical mass of $\sim 4\times 10^8$~M$_\odot$. Now with a
sample of up to 523 members, we will analyze how this ratio varies
when exploring beyond the immediate Solar vicinity.

\begin{figure}
   \centering
   \includegraphics[width=\hsize]{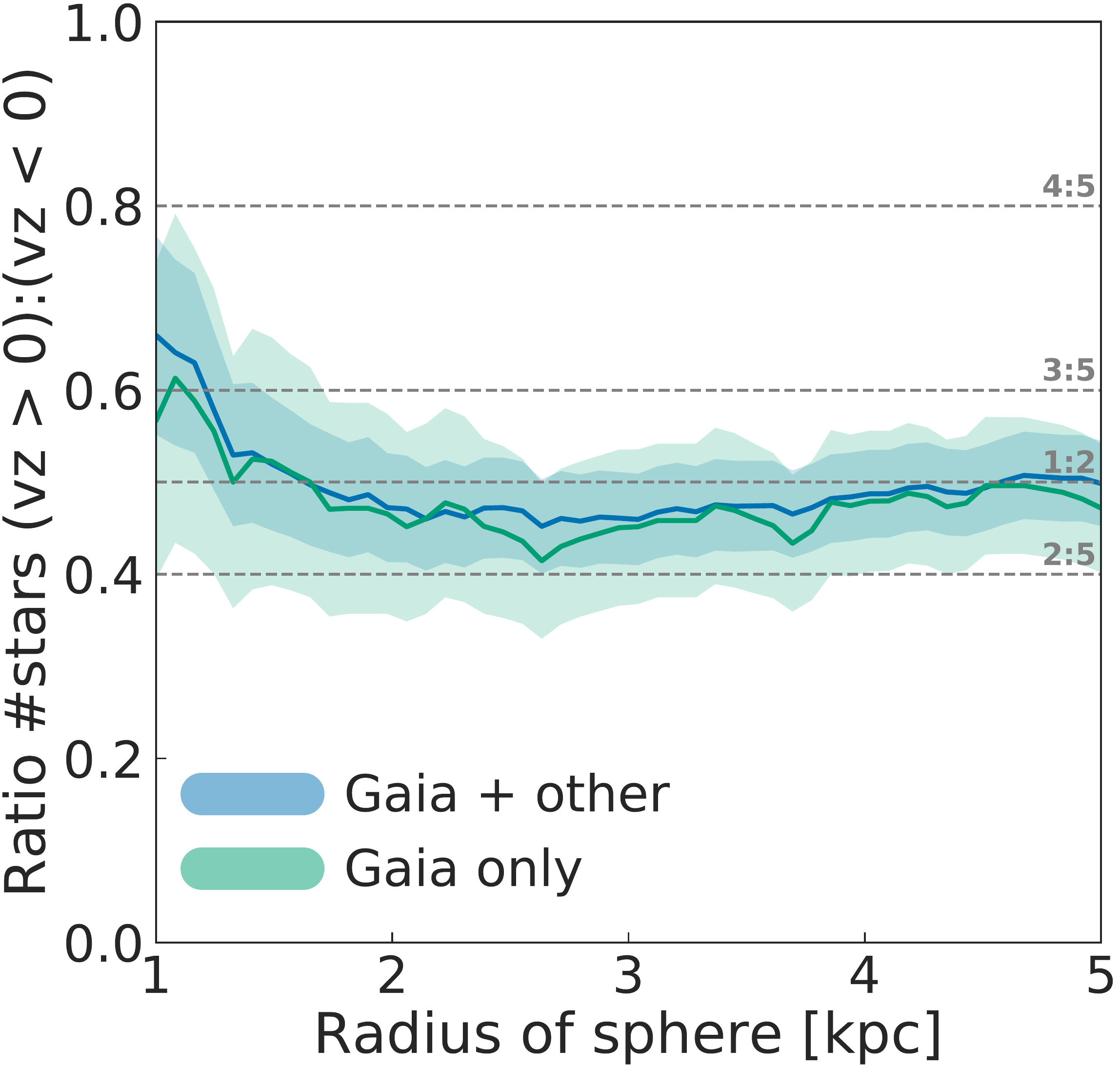}
   \caption{Ratio of the number of stars in the two clumps in $V_Z$
     using selection box B, for different volumes and for the two
     samples of stars as indicated. The shaded area corresponds to the
     Poisson error on the measured ratio.}
         \label{fig:ratios}
\end{figure}

Figure \ref{fig:ratios} shows the ratio of the number of stars in the
two streams in $V_Z$ for selection B, as a function of the extent of
the volume considered. Blue indicates the ratio of all the stars and
green is for the 6D {\it Gaia}-only sample. The central lines plot the
measured ratio and the shaded areas correspond to the 1$\sigma$ Poissonian
error, showing that there is good agreement between the samples. The
dashed lines at 1:2, 3:5 and 2:5 encompass roughly the mean and the
scatter in the ratio. Evidently, the ratio drops slightly beyond the
$1$~kpc volume around the Sun, however, overall it stays rather
constant and takes a value of approximately 1:2 for stars with
$V_Z>0$ relative to those with $V_Z<0$. 

\subsection{HR diagram and metallicity information}\label{sec:phot}

\begin{figure}
   \centering
   \includegraphics[width=\hsize]{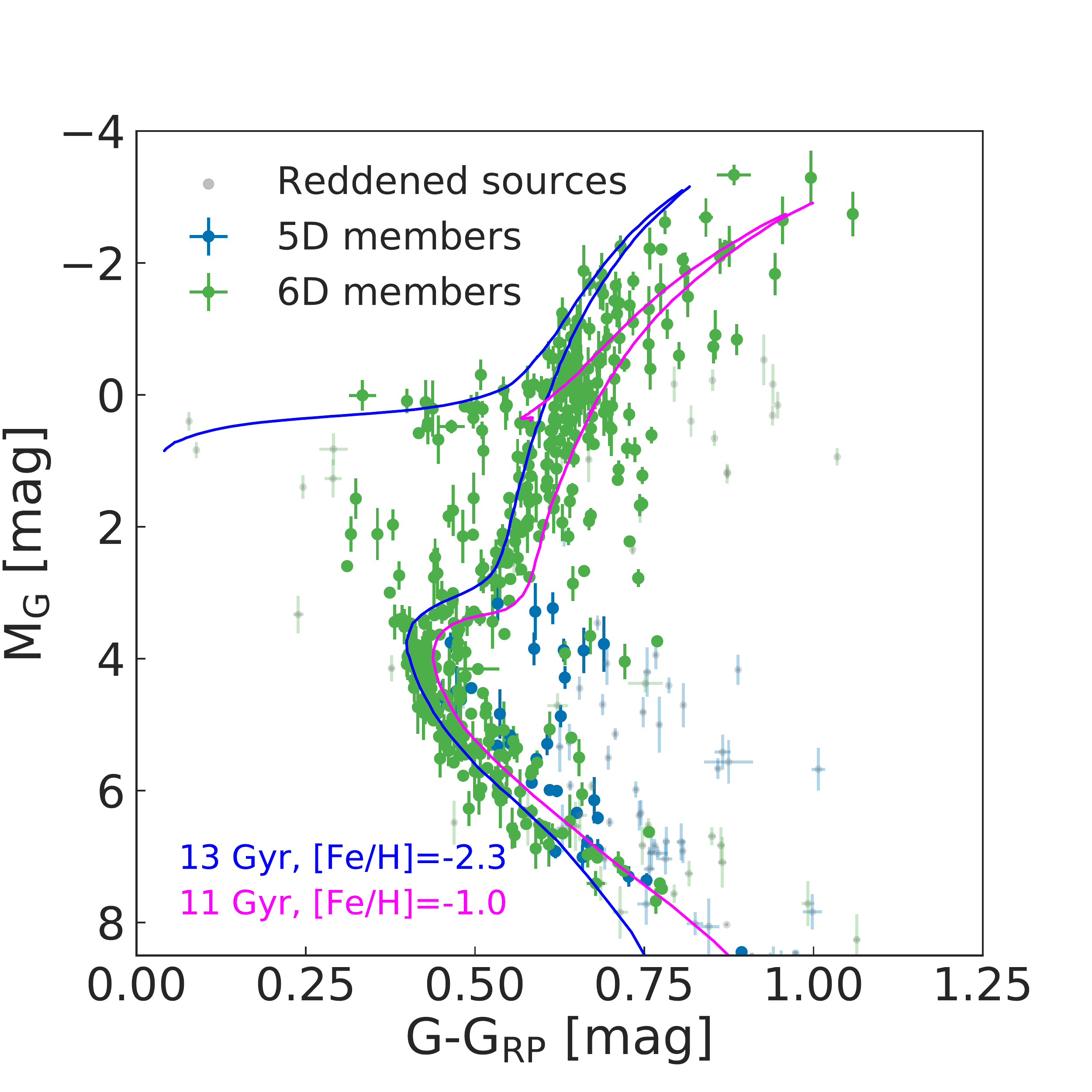}
   \caption{Hertzsprung-Russel diagram of members of the Helmi
     streams. Those identified in the 6D sample are shown in green,
     while those without radial velocities are indicated with blue
     symbols. Members that are likely highly reddened are indicated in
     light grey. Superimposed are single population isochrones taken
     from \cite{Marigo2017}. They serve to illustrate that the Helmi
     streams include a range of old, metal-poor stellar populations
     which did not form in a single event.}
         \label{fig:HRdiagram}
\end{figure}

Photometry from {\it Gaia} combined with auxiliary metallicity
information from the APOGEE/RAVE/LAMOST surveys can give us insights
into the stellar populations of the Helmi streams. By using the {\it
  Gaia} parallaxes we construct the Hertzsprung-Russell (HR) diagram
shown in Fig.~\ref{fig:HRdiagram}. We have used here a high
photometric quality sample by applying the selection criteria from
\cite{Arenou2018} and described in Section \ref{sec:5Dmembers}. Note that
since the photometry in the BP passband is subject to some systematic
effects especially for stars in crowded regions
\citep[see][]{GaiaCollaboration2018brown}, we use ($G - G_\mathrm{RP}$)
colour.

In Fig.~\ref{fig:HRdiagram} the stream members from the 6D sample are
plotted with green symbols, and in blue if they are from the 5D
dataset. On the basis of a colour-colour diagram we have identified
stars that are likely reddened by extinction, and these are indicated
with light grey dots. Since most stars follow a well-defined sequence
in the $[(G - G_\mathrm{RP}),(G - G_\mathrm{BP})]$ space, outliers can
be picked out easily. We consider as outliers those stars with a
$(G - G_\mathrm{BP})$ offset greater than $0.017$ from the sequence
(i.e. 5$\times$ the mean error in the colours used). We find that
especially the members found in the 5D dataset appear to be
reddened. This is expected as all of these stars are located at low
Galactic latitude (within 15 degrees from the galactic centre or
anti-centre). Fig.~\ref{fig:HRdiagram} shows also that we are biased towards
finding relatively more intrinsically bright than fainter stars, and this is due to
the quality cuts applied and to the magnitude limits of the samples
used. We expect however that there should be many more fainter, lower
main sequence stars that also belong to the Helmi streams, hidden in
the local stellar halo.

The HR diagram shown in Fig.~\ref{fig:HRdiagram} does not resemble
that of a single stellar population, but rather favours a wide stellar
age distribution of $\sim 2$ Gyr spread, based on the width of the
main sequence turn-off. To illustrate this we have overlayed two
isochrones from \cite{Marigo2017} for single stellar populations of 11
and 13 Gyr old age and with metallicities [Fe/H]~$=-1.0$ and $-2.3$
respectively. Note that to take into account the difference
between the theoretical and actual {\it Gaia} passbands
\citep{Weiler2018RevisedPassbands}, we have recalibrated the
isochrones on globular clusters with similar age and metallicity
\citep[NGC104, NGC6121, NGC7099, see][]{Harris1996AWAY}, which led to
a shift in ($G - G_\mathrm{RP}$) colour of 0.04 mag.

\begin{figure}
   \centering
   \includegraphics[width=\hsize]{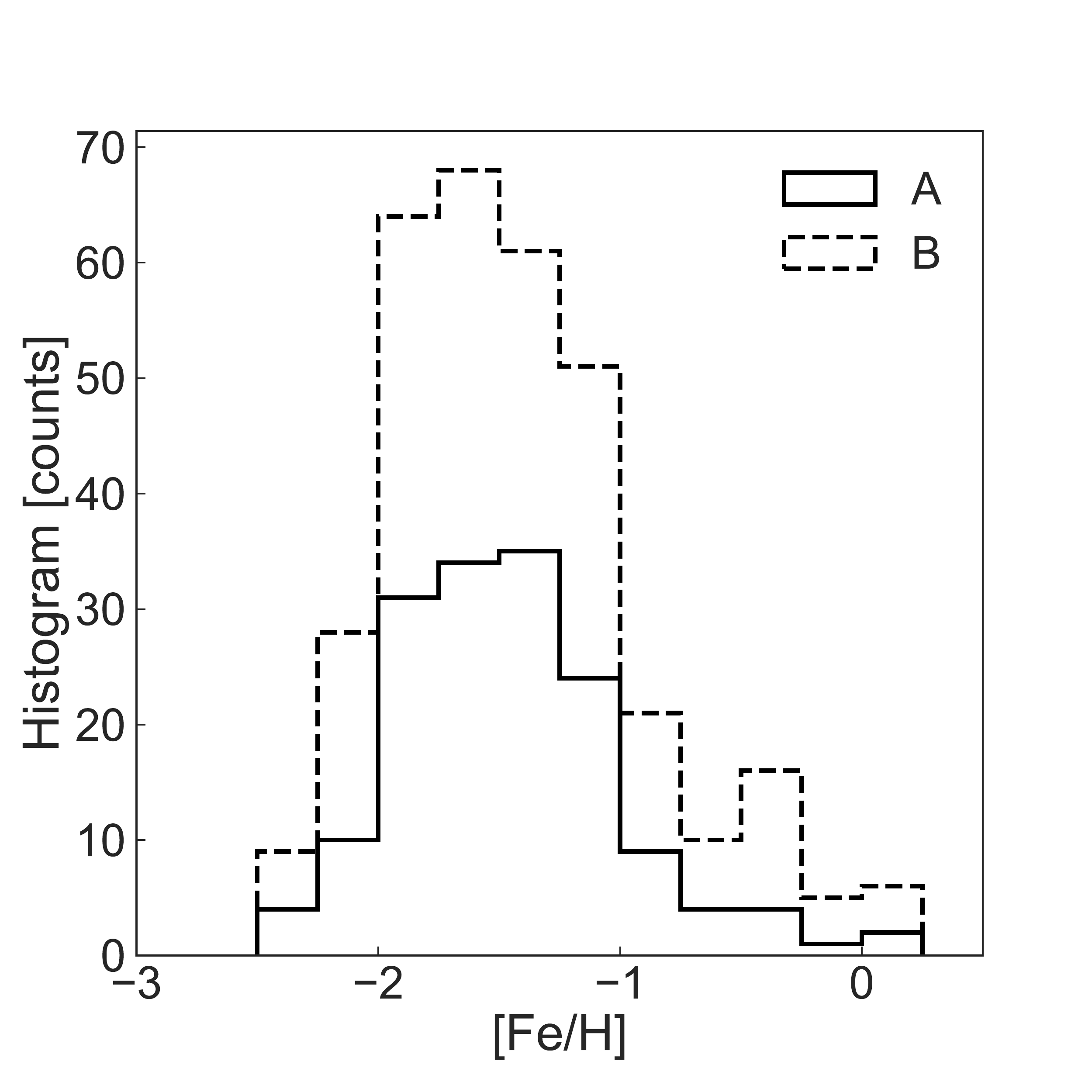}
   \caption{Histogram of the metallicities of the Helmi streams stars
     that are in the APOGEE/RAVE/LAMOST datasets. The distributions
     are very similar for selections A \& B, peaking at
     [Fe/H]~$\sim -1.5$ and revealing a broad range
     of metallicities for the Helmi streams stars. The metal-rich tail
     ([Fe/H] $\sim$ $-0.5$) is likely due to contamination from the
     thick disk but it is minimal for both selection boxes.}
         \label{fig:metallicity}
\end{figure}

The spread in metallicities used for the isochrones is motivated by
the metallicity distribution shown in Figure~\ref{fig:metallicity}, and derived using the
stream members found in the APOGEE/RAVE/LAMOST datasets. We have used
the following metallicity estimates for the different surveys: {\tt
  Met\_K} for RAVE, {\tt FE\_H} for APOGEE and {\tt feh} for
LAMOST. The distribution plotted in Fig.~\ref{fig:metallicity} shows a
range of metallicities [Fe/H]~$ = [-2.3,-1.0]$, with a peak at
[Fe/H]~$ = -1.5$. The small tail seen towards the metal-rich end,
i.e. at [Fe/H]~$\sim -0.5$ is likely caused by contamination from the
thick disk. The shape of the distribution shown in Figure
\ref{fig:metallicity} is reminiscent of that reported by
\cite{Klement2009a} and by \cite{Smith2009} for much smaller samples
of members of the Helmi streams. \cite{Roederer2010CHARACTERIZING}
have carried out detailed abundance analysis of the original stream
members which confirm the range of $\sim 1\,\mathrm{dex}$ in [Fe/H]
found here. All this evidence corroborates that the streams originate
in an object that had an extended star formation history.

%###################################################################
%-------------------------------------------------------------------
%###################################################################

\section{Simulating the streams} \label{sec:simulations} 

We focus here on N-body experiments we have carried out to reproduce
some of the properties of the Helmi streams. To this end, we used a
modified version of {\tt GADGET 2}
\citep{Springel2005SimulationsQuasars} that includes the host rigid,
static potential described in section Section \ref{sec:beyondcore} to model
the Milky Way.

The analysis presented in previous sections, and in particular the HR
diagram and metallicity distribution of member stars, supports the
hypothesis that the Helmi streams stem from a disrupted (dwarf)
galaxy. We therefore model the progenitor of the streams as a dwarf
galaxy with a stellar and a dark matter component. We consider four
possible progenitors whose characteristics are listed in Table
\ref{tabel:dwarfs_specs}.

For the stellar component, we use $10^5$ particles distributed
following a Hernquist profile, whose structural properties are
motivated by the scaling relations observed for dwarf spheroidal
galaxies \citep{Tolstoy2009Star-FormationGroup}. For the dark matter
halo we use $6 \times 10^5$ particles following a truncated NFW
profile \citep[similar the model introduced in][but where the truncation radius $\text{r}_{\textrm{c,trunc}}$ and the decay radius $\text{r}_{\textrm{d,trunc}}$ are specified independently]{Springel1999TidalCosmologies}, with
characteristic parameters taken from \cite{Correa2015TheRelation}. We
truncate the NFW halo at a radius where its average density is three
times that of the host (at the orbital pericentre). After setting the
system up using the methods described in \cite{Hernquist1993}, we let
it relax for 5 Gyr in isolation. We then place it on an orbit
around the Milky Way. This orbit is defined by the mean position and
velocity of the stars that were identified as core members of the
stream with $V_Z < 0$\footnote{We take the $V_Z < 0$ clump 
  as this has the largest number of members.}.

\begin{table}
  \caption{Structural parameters of the simulated dwarf galaxies. Note that we quote both the dark halo's original and truncated mass. This truncation depends on two parameters: $\text{r}_{\textrm{c,trunc}}$  and $\text{r}_{\textrm{d,trunc}}$, the cut-off and the decay radii, respectively.}
\label{tabel:dwarfs_specs}
\centering
\begin{tabular}{c c c c c}
\hline\hline
 & prog. 1 & prog. 2 & prog. 3 & prog. 4 \\
\hline      
$\text{M}_{*}$ ($\text{M}_{\odot}$) & $5\cdot 10^6$ & $10^7$ & $5\cdot 10^7$ & $10^8$ \\
$\text{r}_{s,*}$ (kpc) & 0.164 & 0.207 & 0.414 & 0.585 \\
$\text{r}_{\text{s,NFW}}$ (kpc) & 1.32 & 1.72 &  3.42 &  4.26 \\
$\text{M}_{\textrm{dm}}$ ($\text{M}_{\odot}$) & $5\cdot 10^8$ & $10^9$ & $5\cdot 10^9$ & $10^{10}$ \\ 
$\text{M}_{\textrm{dm,trunc}}$ ($\text{M}_{\odot}$) & $1.9\cdot 10^8$ & $3.8\cdot 10^8$ & $1.84\cdot 10^9$ & $3.62\cdot 10^9$ \\ 
$\text{r}_{\textrm{c,trunc}}$ (kpc) & 2.0 & 2.5 & 4.1 & 5 \\
$\text{r}_{\textrm{d,trunc}}$ (kpc) & 0.65 & 0.86 & 1.62 & 2.13 \\
\hline
\end{tabular}
\end{table}

\begin{figure*}
   \centering
   \includegraphics[width=\hsize]{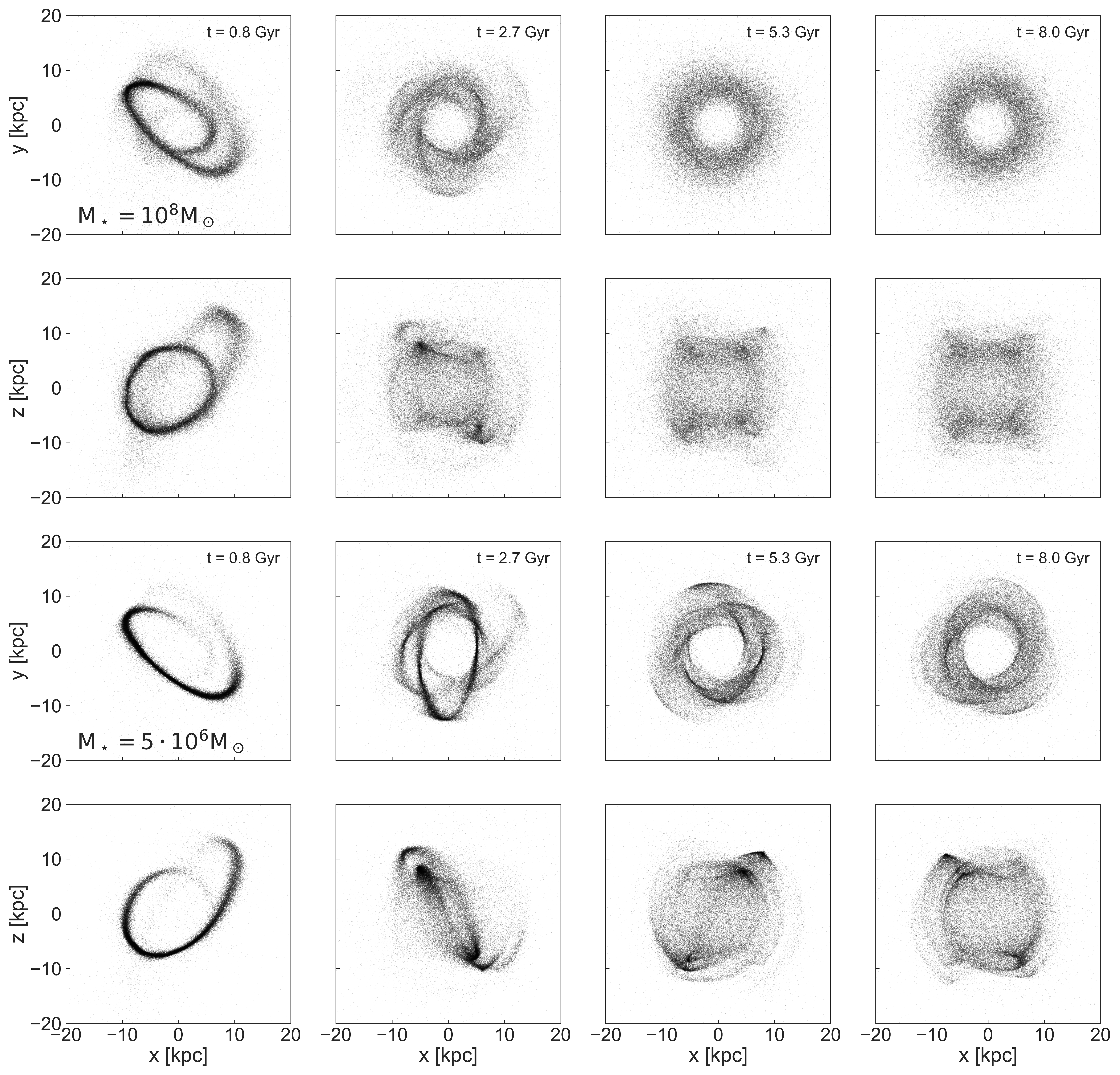}
   \caption{Spatial evolution of the Helmi streams at 
     four different snapshots in our simulations. The top two rows
     show the evolution of a progenitor with a stellar mass of
     $10^{8}\mathrm{M}_\odot$, while the bottom two rows correspond to
     a system with a stellar mass of $5\times
     10^{6}\mathrm{M}_\odot$. In both cases, the orbit of the
     progenitor is the same. The appearance of the debris
     is seen to depend on the time since accretion as well as on the mass of the
     progenitor.}
         \label{fig:simulations_evolution}
\end{figure*}

\begin{figure}
  \centering
  \includegraphics[width=\hsize]{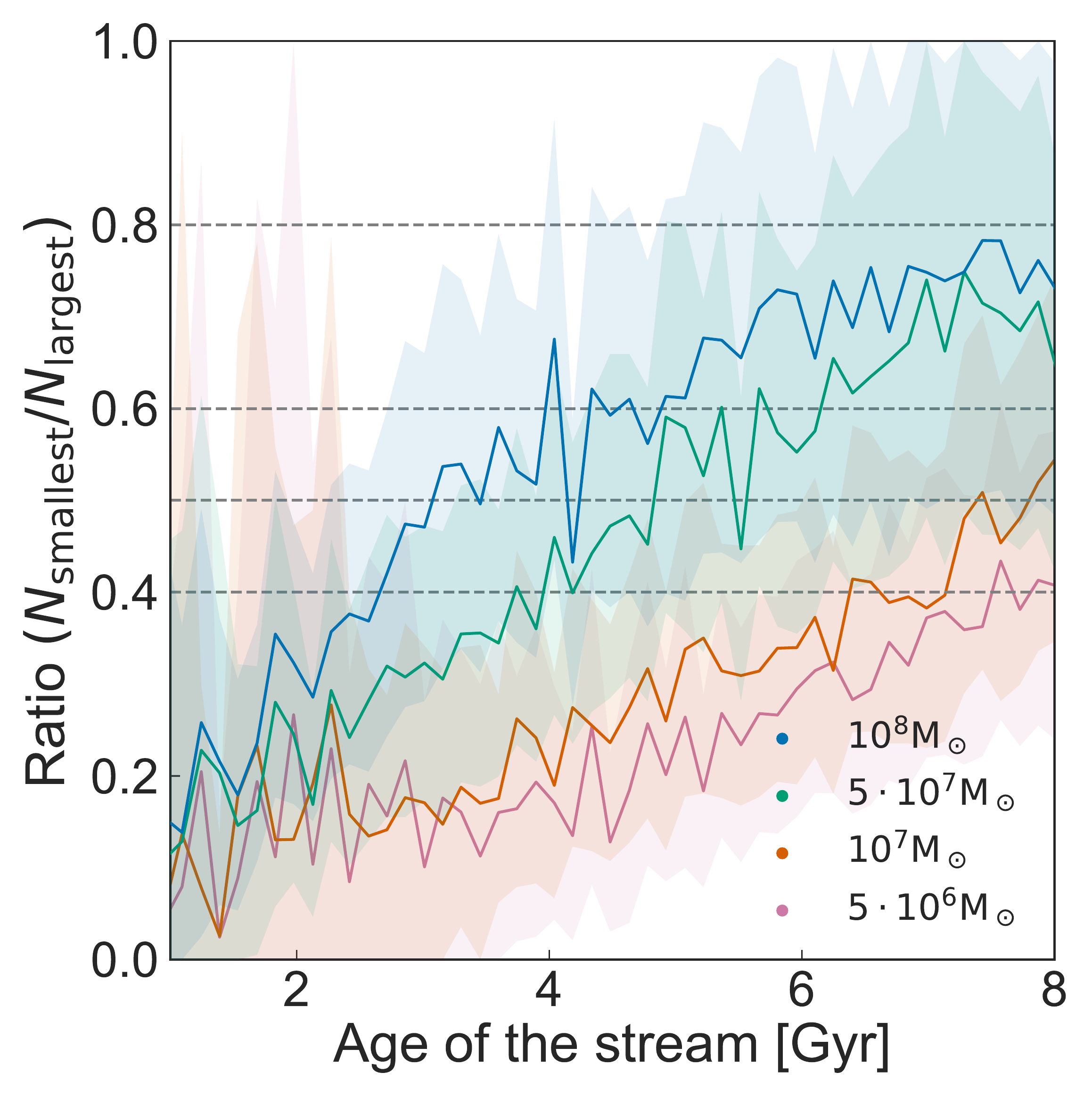}
      \caption{Ratio of the numbers of stars in the less populated $V_Z$
    clump compared to that of the more populated clump, as a function of
    time for our N-body simulations. The solid lines show the mean
    values, averaged over 25 different volumes spread uniformly along
    a circle of 1 kpc at the Solar radius.}
         \label{fig:progprops_sim}
\end{figure}

\begin{figure}
  \centering
  \includegraphics[width=\hsize]{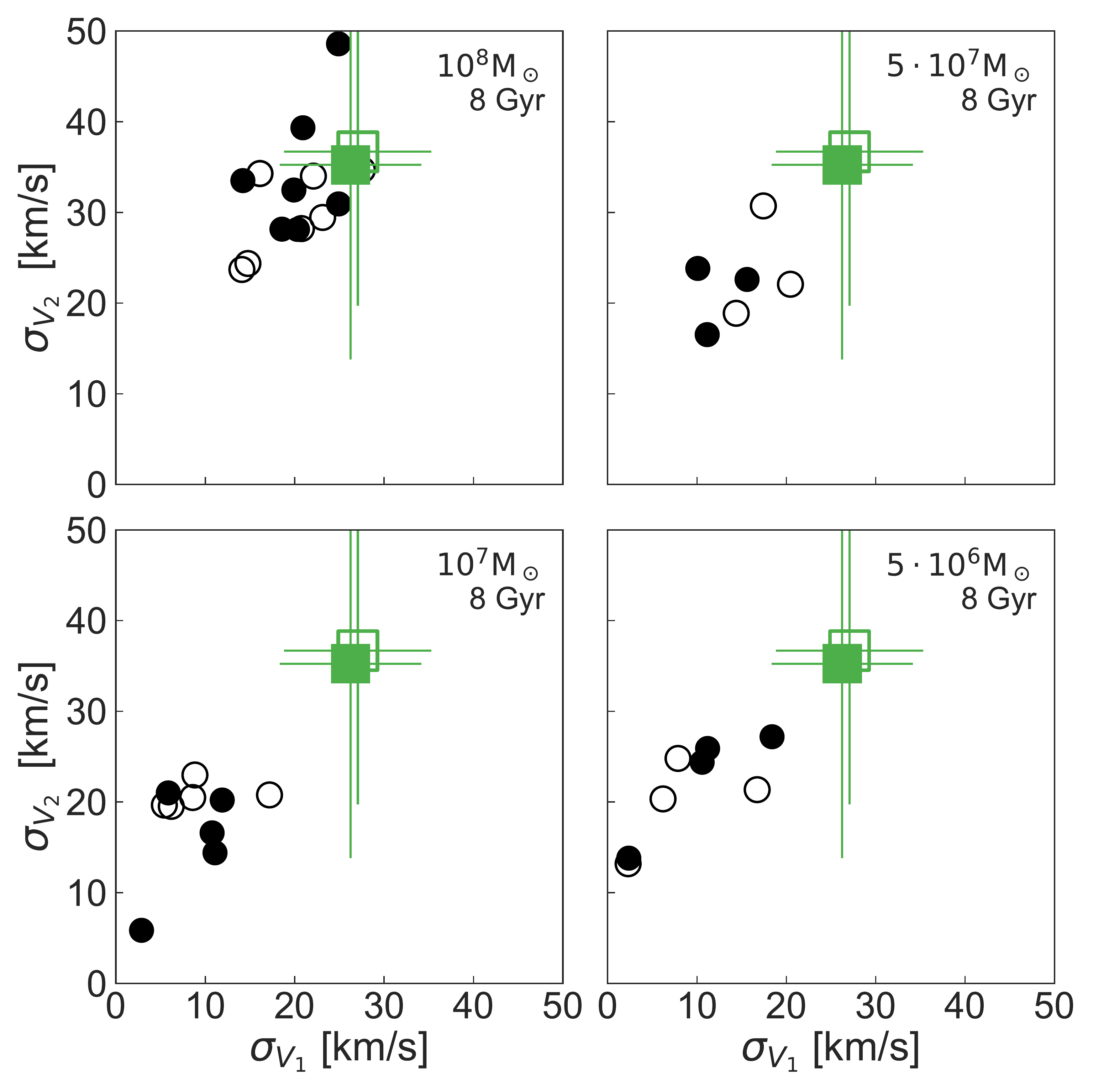}
  \caption{Velocity dispersions of the simulated streams 8 Gyr after
    accretion. These have been computed using the
    principal axis of the velocity ellipsoid of stars in a given $V_Z$
    clump, open/closed markers are used for the smallest/largest clump
    respectively. We show the results for volumes satisfying the ratio
    in the number of stars in the clumps as observed for the data. The
    green markers indicate the measured velocity dispersion from the
    data, the error bars illustrate the scatter in 1000 randomly
    downsampled sets of the streams members.}
         \label{fig:vel_disp}
\end{figure}

Figure \ref{fig:simulations_evolution} illustrates the evolution of a
high mass dwarf galaxy in the top two rows, and for a low mass dwarf
galaxy in the bottom two rows. In each panel, we show the star
particles in galactocentric Cartesian coordinates for different times
up to 8 Gyr after infall. This comparison shows that increasing the
mass results in more diffuse debris. On the other hand, the time since
accretion has an impact on the length of the streams and on how many
times the debris wraps around the Milky Way.

\subsection{Estimation of the mass and time of accretion} \label{sec:progprop}

To constrain the history of the progenitor of the Helmi streams we use
the ratio of the number of stars in the two clumps in $V_Z$ as well as
their velocity dispersion. Typically, the simulated streams do not
have a uniform spatial distribution and they also show variations on
small scales. Furthermore, the azimuthal location of the Sun in the
simulations is arbitrary. Therefore, and also to even out some of the
small-scale variations, we measure the ratio of the number of stars in
the streams and their velocity dispersion in 25 volumes of 1 kpc
radius located at 8.2 kpc distance from the centre, and distributed
uniformly in the azimuthal angle $\phi$.

Fig.~\ref{fig:progprops_sim} shows the mean of the ratio for the
different volumes as a function of time with solid curves, with the
colours marking the different progenitors listed in
Table~\ref{tabel:dwarfs_specs}. The shaded areas correspond to the
mean Poissonian error in the measured ratio for the different
volumes. The horizontal dashed lines are included for guidance and
correspond to the lines shown in Fig.~\ref{fig:ratios} for the actual
data.

The mean ratio is clearly correlated with the properties of the progenitor,
the most massive one being first to produce multiple streams in a
given volume. Massive satellites have a larger size and velocity
dispersion, which causes them to phase mix more quickly because of the
large range of orbital properties (energies, frequencies). Based
solely on Fig.~\ref{fig:progprops_sim}, and taking a ratio between 0.55 and 0.7 as
found using the {\it Gaia}-only sample in a 1~kpc sphere, we would
claim that there is a range of possible ages of the stream, with the
youngest being $\sim 4.5$ Gyr for the most massive progenitor, while
for the lowest mass object the age would have to be at least 8 Gyr.

\begin{figure*}
   \centering
   \includegraphics[width=\hsize]{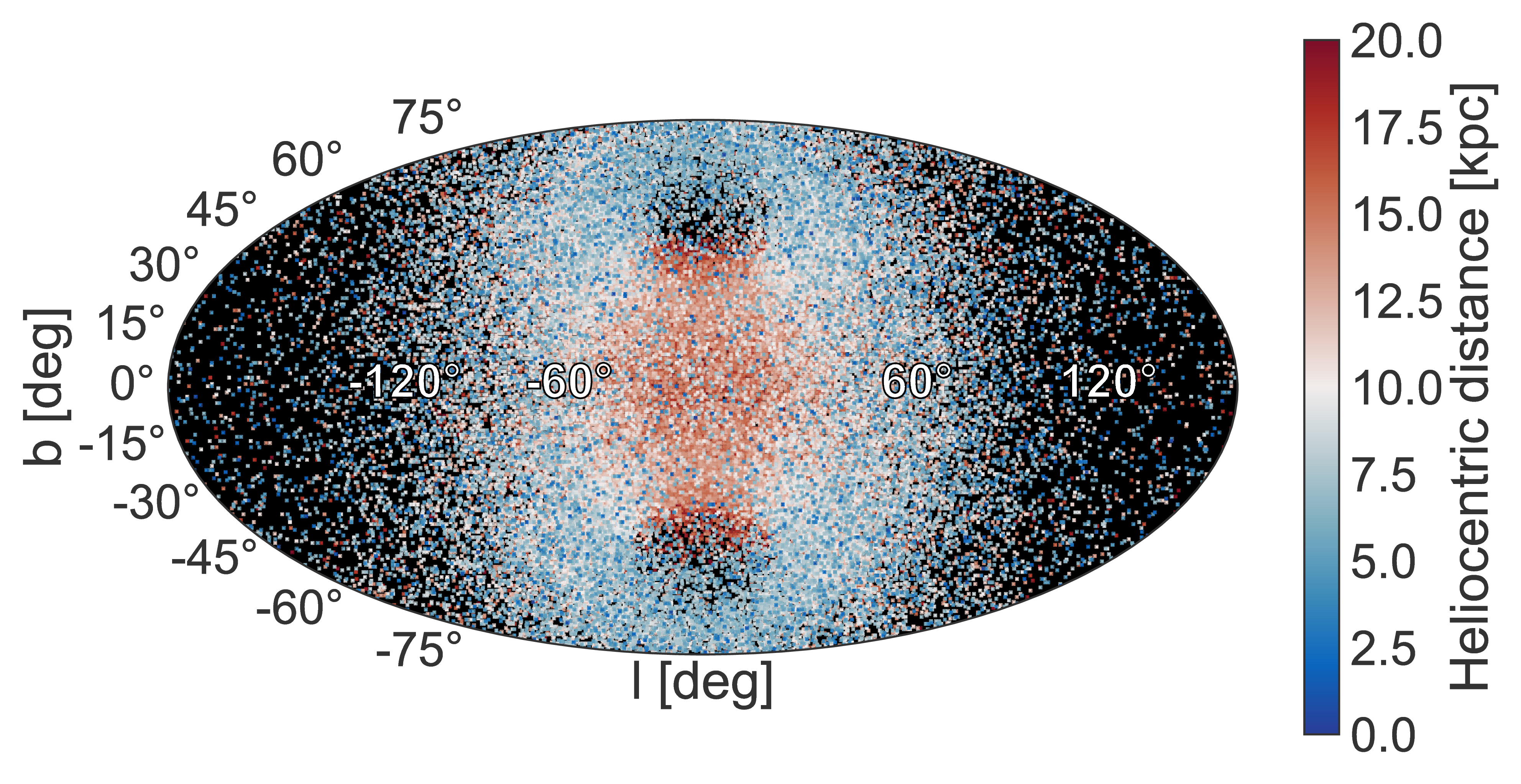}
   \caption{Mollweide projection in
     Galactic sky coordinates $l$ and $b$ of the distribution of star
     particles for the simulation of a progenitor dwarf galaxy with a
     stellar mass of $10^{8}\,\mathrm{M}_\odot$. The colour-coding
     gives the heliocentric distance of the particles.}
         \label{fig:simulations_onsky}
\end{figure*}

A different way of probing the properties of the progenitor is to use
the velocity dispersion of the streams. In Fig.~\ref{fig:vel_disp} we
show the dispersion along two principal axes of the velocity ellipsoid
for the 1~kpc volumes that satisfy the ratio-constraint on the number
of stars in the two $V_Z$ clumps, with open/closed markers used for
the least/most populated clump. Green markers indicate the measured
velocity dispersions. Typically the simulations have fewer particles
in such 1 kpc volumes than observed in the debris. Therefore, we
down-sample the data to only 15 stars per stream, which corresponds to
the average number of particles in the simulations. The error bars
indicate the maximum scatter in the velocity dispersion for 1000
such random downsampled sets.

The ellipsoid is roughly aligned in cylindrical coordinates with
$V_1$, $V_2$, and $V_3$ corresponding respectively to
$V_\phi,\,V_Z,\,V_R$.  Note that only for the most massive progenitor
(and for times > 5 Gyr) the velocity dispersions are in good agreement
with the observed values for $\sigma_{V1}$ and $\sigma_{V2}$, and that
in all the remaining simulations, the dispersions are too small
compared to the data. We should point out however, that the largest
velocity dispersion (i.e. that along $V_R$) is less well reproduced in
our simulations, possibly indicating that the range of energies of the
debris is larger. However, if we clip 1$\sigma$ outliers in $V_R$ for
the data, and recompute $\sigma_{V3}$, we find much better agreement
with the simulations (while the values of $\sigma_{V1}$ and
$\sigma_{V2}$ remain largely the same). The clipped members of the
Helmi streams are all in the high energy, $E$, tail (see
Fig.\ref{fig:IOMselection}), suggesting perhaps that the object has suffered
some amount of dynamical friction in its evolution.

Combining the information of the ratio (Fig.~\ref{fig:progprops_sim})
and the corresponding velocity dispersion measurements (
Fig.~\ref{fig:vel_disp}) we therefore conclude that the most likely
progenitor of the Helmi streams was a massive dwarf galaxy with a
stellar mass of $\sim 10^{8}\,\mathrm{M}_\odot$. In general, the
simulations suggest a range of plausible accretion times from $5 - 8$
Gyr. Although these estimates of the time of accretion are different
than those obtained by \cite{Kepley2007HALONEIGHBORHOOD}, our results
are consistent when a similar progenitor mass is used (i.e. of models
1/2), which however, does not reproduce well the observed kinematical
properties of the streams.

\subsection{Finding new members across the Milky Way}\label{sec:newmembers}

To investigate where to find new members of the Helmi streams beyond
the Solar neighbourhood we turn to the simulations.

In Fig.~\ref{fig:simulations_onsky} we show a sky density map of stars
from the N-body simulation corresponding to the dwarf galaxy with $M_* = 10^{8}\,\mathrm{M}_\odot$, 8.0~Gyr after
accretion. The coordinates shown here are galactic $l$ and $b$ plotted
in a Mollweide projection with the Galactic centre located in the
middle (at $l=0$). Note that nearby stars (in blue) are mainly
distributed along a ``polar ring-like'' structure between
longitudes $\pm 60^\circ$ (see for comparison
Fig.~\ref{fig:spatial_distribution_XZ}). The most distant members are
found behind the bulge (in red), but because of their location they
may be difficult to observe.

\section{Association with globular clusters}\label{sec:globassoc}

If the progenitor of the Helmi streams was truly a large dwarf galaxy,
it likely had its own population of globular clusters
\citep[see][]{Leaman2013,DiederikKruijssen2018ThePopulation}. To this
end we look at the distribution of the debris in IOM-space for the
simulation of the progenitor with $M_* = 10^{8}\,\mathrm{M}_\odot$ and
overlay the data for the globular clusters from
\cite{GaiaCollaboration2018Helmi}.

\begin{figure}
   \centering
   \includegraphics[width=\hsize]{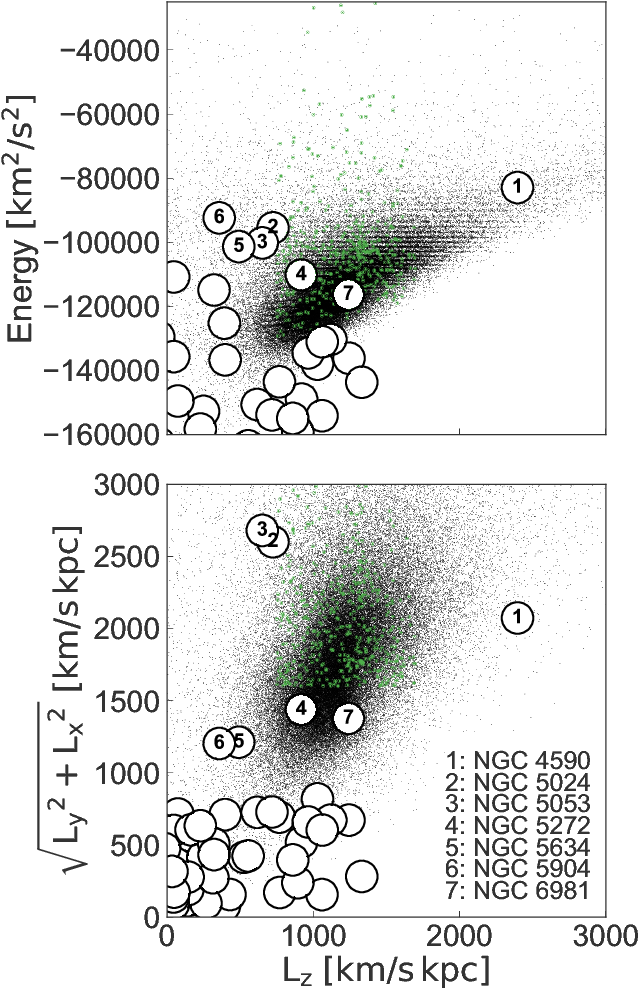}
   \caption{$E-L_z$ (top) and the
     $L_\perp - L_z$ (bottom) distributions of the simulated stars (in
     black), together with the selected 6D stream members (in
     green). Plotted here are the star particles from the simulation
     of the most massive dwarf, with a stellar mass of
     $10^{8}\,\mathrm{M}_\odot$ and accretion time of 8.0 Gyr. Large
     symbols indicate the location of Milky Way globular
     clusters. Those possibly associated with the Helmi streams based
     on their location in these panels are numbered.}
         \label{fig:GCfig}
\end{figure}

Figure \ref{fig:GCfig} shows the energy $E$ vs $L_z$ (top) and the
$L_\perp$ vs $L_z$ (bottom) of star particles in the simulation
(black) and the stream members (green) together with the globular
clusters (white open circles). We have labelled here the globular
clusters that show overlap with the streams members in this
space. Those that could tentatively be associated on the basis of
their orbital properties are: NGC 4590, NGC 5024, NGC 5053, NGC 5272,
NGC 5634, NGC 5904, and NGC 6981. 

\begin{figure}
   \centering
   \includegraphics[width=\hsize]{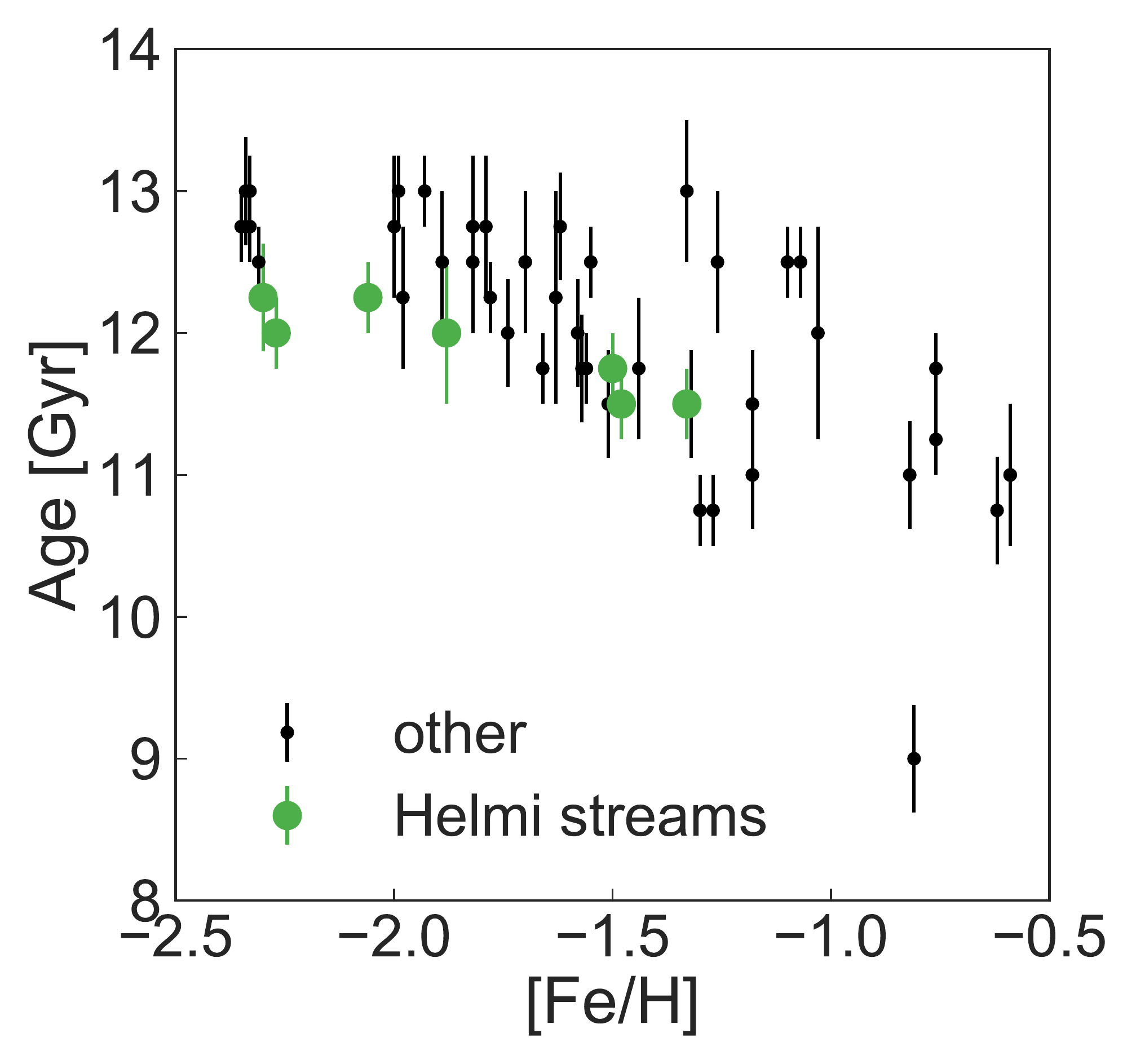}
      \caption{Age-metallicity distribution of Milky Way globular clusters based on \cite{Vandenberg2013METHODISSUES}. The green symbols mark the clusters associated with the Helmi streams on the basis of their orbital properties. They follow a well-defined age-metallicity relation.}
         \label{fig:age-met}
\end{figure}

This set of globular clusters shows a moderate range in age and
metallicity: they are all old with ages $\sim 11-12$ Gyr and
metal-poor with metallicities [Fe/H] = [$-2.3$,$-1.5$]. These age
estimates are from \cite{Vandenberg2013METHODISSUES}, while for NGC
5634 we set it to $12$~Gyr from comparison to NGC 4590 based on the
zero-age HB magnitude \citep{Bellazzini}, and assume an uncertainty of
0.5~Gyr. Interestingly, Fig.~\ref{fig:age-met} shows that the clusters
follow a relatively tight age-metallicity relation, and which is
similar to that expected if they originate in a progenitor galaxy of
$M_* \sim 10^{7} - 10^8 \,\mathrm{M}_\odot$
\citep[see][Fig.~4]{Leaman2013}.

\begin{figure}
   \centering
   \includegraphics[width=\hsize]{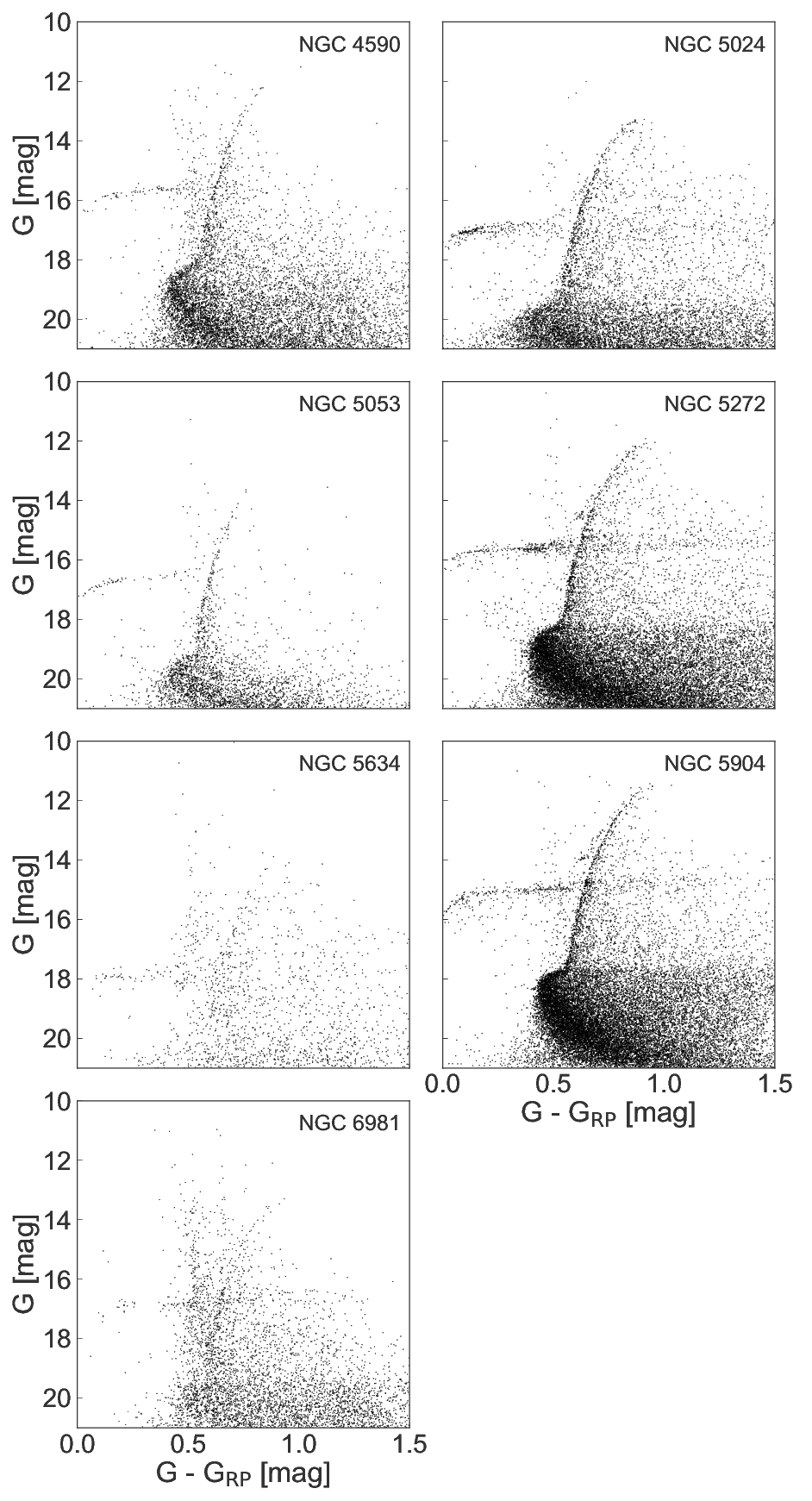}
      \caption{{\it Gaia} colour-magnitude diagrams for the globular clusters that are likely associated with the Helmi streams on the basis of their orbital properties. All the CMDs correspond to stellar populations that are old and metal-poor.}
         \label{fig:GCCMD}
\end{figure}

Fig.~\ref{fig:GCCMD} shows the {\it Gaia} colour-magnitude diagrams
(CMD) of the globular clusters tentatively associated with the Helmi
streams. Although not all CMDs are well-populated because of
limitations of the {\it Gaia} DR2 data, their properties do seem to be
quite similar, increasing even further the likelihood of their
association to the progenitor of the Helmi streams.

Some of the associated globular clusters, namely NGC 5272,
NGC 5904, and NGC 6981 have in fact, been suggested to have an
accretion origin \citep[of a yet unknown progenitor, see][and references
therein]{DiederikKruijssen2018ThePopulation}. Two other clusters, NGC
5024 and NGC 5053, have at some point been linked to
Sagittarius, although recent proper motion measurements have
demonstrated this association is unlikely
\citep[see][]{Law2010ASSESSINGGALAXY,Sohn2018AbsoluteMass,
  GaiaCollaboration2018Helmi}. Also NGC 5634 has been related to
Sagittarius \citep{Law2010ASSESSINGGALAXY,Carretta2017ChemicalGalaxy}
based on its position and radial velocity. However, the proper motion
of the system measured by \cite{GaiaCollaboration2018Helmi} is very
different from the prediction by for example, the
\cite{Law2010ASSESSINGGALAXY} model of the Sagittarius streams.

%\begin{table}
%\caption{Globular cluster parameters compiled from two external catalogues.}
%\label{tabel:globularclusters}
%\centering
%\begin{tabular}{c c c c}
%\hline\hline
%NGC & [Fe/H] & Age & reference \\
%\hline      
%4590 & -2.27 & 12.00 & \citep{Vandenberg2013METHODISSUES}\\
%5024 & -2.06 & 12.25 & idem \\
%5053 & -2.30 & 12.25 & idem\\
%5272 & -1.50 & 11.75 & idem\\
%5634 & -1.88 &  - & \citep[2010 edition]{Harris1996AWAY}\\
%5904 & -1.33 & 11.50 & \citep{Vandenberg2013METHODISSUES}\\
%6981 & -1.48 & 11.50 & idem\\
%\hline
%\end{tabular}
%\end{table}

\section{Conclusions}\label{sec:conclusion}

Using the latest data from {\it Gaia} DR2 combined with the
APOGEE/RAVE/LAMOST surveys we find hundreds of new tentative members
of the Helmi streams. In the 6D sample that we built, we identified
523 members on the basis of their orbital properties, in particular
their energy and angular momenta. On the other hand, we found 105 stars
in the full {\it Gaia} 5D dataset in two 15$^{\rm o}$-radius fields around
the galactic centre and anticentre, using only the tangential
velocities of the stars (which translate directly into two components
of their angular momenta). Despite the large number of newly
identified members we expect that many, especially faint stars are
still hiding, even within a volume of 1 kpc around the Sun.

Having such an unprecedented sample of members of the streams allows
us to characterize the streams and the nature of their progenitor. The
HR diagram of the members suggests an age range of $\sim 11-13$ Gyr,
while their metallicity distribution goes from [Fe/H]~$\sim -2.3$ to
$-1.0$, with a peak at [Fe/H]$\sim -1.5$. We are also able to
associate to the streams seven globular clusters on the basis of their
dynamical properties. These clusters have similar ages and
metallicities as the stars in the streams. Remarkably they follow a
well-defined age-metallicity relationship, and similar to that
expected for clusters originating in a progenitor galaxy of
$ M_* \sim 10^7 - 10^8 \,\mathrm{M}_\odot$ \citep{Leaman2013}.

This relatively high value of the stellar mass is also what results
from N-body simulations that aim to recreate the observed dynamical
properties of the streams. From the ratio of the number of stars in
the two clumps in $V_Z$ and their velocity dispersion we estimate the
time of accretion to be in the range $5 - 8$~Gyr and a stellar mass
for the dwarf galaxy of $\sim 10^8\mathrm{M}_\odot$.

Although 5 Gyr ago would imply a relatively recent accretion event,
one might argue that the object was probably on a less bound orbit and
sunk in via dynamical friction (thanks to its large mass) and started
to get disrupted then. This could explain the mismatch between
the age of the youngest stars in the streams (approximately 11 Gyr old), and
the time derived dynamically.

Despite the fact that the simulations are able to recreate the
observations reasonably well, they fail to reproduce fully the
observed velocity distribution in particular in the radial
direction. This could be due to the lack of dynamical friction, but
also by the limited exploration of models for the potential of the
Milky Way. Other important improvements will be to consider the
inclusion of gas particles and star formation in the simulations, as
well as different initial morphologies for the progenitor systems (not
only spherical, but also disk-like).

Originally, H99 determined that 10\% of the stellar halo mass beyond
the Solar radius could belong to the progenitor of the Helmi
streams. The lack of a significant increase in the number of members
subsequently discovered by other groups \citep{Chiba2000}, led to the
suggestion that the fraction may be lower. Our best estimate of the
stellar mass of the progenitor of the Helmi stream is
$\sim 10^8\,\mathrm{M}_\odot$, implying that it does significantly
contribute to the stellar halo. For example \cite{Bell2008THEHALO}
estimate a stellar mass for the halo of
$(3.7\pm1.2)\times10^{8}\,\mathrm{M}_\odot$ between galactocentric
radii of 1 to 40 kpc, and hence being a lower limit to the total
stellar halo mass. Other estimates, based on the local density of halo
stars give $7 - 10 \times10^{8}\mathrm{M}_\odot$
\citep[see][]{Morrison1993,Bland-Hawthorn2016TheProperties}. This implies that the
Helmi streams may have contributed $\sim 10 - 14\%$ of the stars
in the Galactic halo.

\begin{acknowledgements}
We gratefully acknowledge financial support from a VICI
grant from the Netherlands Organisation for Scientific Research (NWO) and from
NOVA. This work has made use of data from the European Space Agency (ESA)
mission Gaia ({\tt http://www.cosmos.esa.int/gaia}), processed by the Gaia
Data Processing and Analysis Consortium (DPAC, {\tt http://www.cosmos.esa.
int/web/gaia/dpac/consortium}). Funding for the DPAC has been provided
by national institutions, in particular the institutions participating in the Gaia
Multilateral Agreement. 
\par
We have also made use of data from: (1) the APOGEE survey, which is part of
Sloan Digital Sky Survey IV. SDSS-IV is managed by the Astrophysical
Research Consortium for the Participating Institutions of the SDSS
Collaboration ({\tt http://www.sdss.org}). (2) the RAVE survey ({\tt
  http://www.rave-survey.org}), whose funding has been provided by
institutions of the RAVE participants and by their national funding
agencies. (3) the LAMOST DR4 dataset, funded by the National
Development and Reform Commission. LAMOST is operated and managed by
the National Astronomical Observatories, Chinese Academy of Sciences.
\par
For the analysis, the following software packages have been used: {\tt vaex} \citep{Breddels2018}, {\tt numpy} \citep{VanDerWalt2011TheComputation}, {\tt matplotlib} \citep{Hunter2007Matplotlib:Environment}
, {\tt jupyter notebooks} \citep{Kluyver2016JupyterWorkflows}.
\end{acknowledgements}

\bibliographystyle{aa} 
\bibliography{helmer} 

\appendix{Appendix: Table of members selection B}

\renewcommand{\thetable}{A\arabic{table}}
\begin{table*}
\caption{{\it Gaia} DR2 source\_ids for selection B}             % title of Table
\label{AppTable:2}      % is used to refer this table in the text
\centering                          % used for centering table
\begin{tabular}{c c c c c}       % centered columns (5 columns)
\hline\hline                 % inserts double horizontal lines
IDs members 1-105 & IDs members 106-210 & IDs members 211-315  & IDs members 316-419 & IDs members 420-523 \\    % table heading 
\hline % inserts single horizontal line
4641558272441088 & 15542185269834624 & 18721418846252288 & 25052342374772736 & 40341188999965056 \\44825306655035520 & 47948469433052800 & 48191667661477120 & 66174455213158656 & 88903903177506304 \\109717108535569920 & 114370963299054720 & 129029789761784320 & 133659416611738496 & 148992625953529216 \\166864465909649024 & 208313266841386496 & 214416419665198976 & 277505056837614336 & 297601582475179520 \\298510126972435200 & 300034840362494592 & 317928636190603136 & 328660281995935232 & 334897296063939712 \\338420371837484160 & 362606329112459136 & 365903386527108864 & 371779211025724032 & 378336075603390208 \\414268497862196224 & 424315422799326464 & 565085174940343296 & 571826104634521856 & 579282859350500480 \\579472799984173824 & 580075916471255040 & 597915355193425152 & 601485847406210688 & 604095572614068352 \\609634774756212480 & 613717261429518336 & 618362698056754816 & 623835615967986560 & 640225833940128256 \\641011469357926144 & 641984640227836288 & 642298482077861888 & 650878108749480960 & 658200993629082240 \\660857241923470336 & 666167814367063296 & 675824721913920768 & 676184743253306752 & 690133560079137920 \\694677944716195712 & 695472307507752320 & 703875080308465536 & 711475458731171712 & 715176071273608704 \\720354564881582080 & 721179847141106816 & 722255002009942528 & 722917384750970752 & 724057994920894592 \\724067027236448128 & 731949048138844800 & 735896535401139072 & 777558680244658176 & 780787744032780544 \\793157490364148224 & 795010747276301952 & 809190053525078272 & 809488433492803072 & 809944799537192320 \\812131247128723712 & 814140020513681024 & 821455518049745280 & 838151499037299328 & 839367249661110912 \\839618793008582272 & 840616123070439552 & 841170654888167808 & 842363247047642368 & 845266198261944960 \\846326471068781824 & 849565494885493888 & 850419781060360448 & 868447614227967744 & 877174369298101504 \\877928256318192512 & 878629435499205632 & 879194790634452352 & 879287699366827520 & 885358240501948032 \\892436552763913344 & 894379527249498496 & 900194908672475648 & 901431176354580480 & 909255575975553920 \\924073900341372160 & 926019211289351424 & 941117327005728000 & 951217299082189696 & 961149020114863872 \\975473938637155968 & 1011601034570885504 & 1021904837208334848 & 1022757474116502400 & 1022993078842418432 \\1033400437435614976 & 1049376272667191552 & 1143069060084735232 & 1155172720304980224 & 1160604689999075456 \\1176187720407158912 & 1179353424836990720 & 1190457117888556160 & 1211628386079715840 & 1243915819907387392 \\1244298759191228800 & 1255012675370113536 & 1258514207587827840 & 1260211058971256320 & 1261193202028905600 \\1262839514533098368 & 1263921674493183488 & 1264302346034442496 & 1275876252107941888 & 1278333454435460096 \\1278999758481584768 & 1286475922158536064 & 1288312793771274880 & 1317046296776785920 & 1319060876956649216 \\1325856580370512768 & 1329198236725539456 & 1351435584519064576 & 1358807878703209472 & 1359836093873456768 \\1361390494077900672 & 1375199947805963520 & 1375468537880511488 & 1376687518318241536 & 1390150935120171776 \\1403917336097206144 & 1415635209471360256 & 1416077522383596160 & 1426065314211359488 & 1445550069004682368 \\1454211574931259008 & 1454848260885706240 & 1459258161504411776 & 1461351734723634944 & 1465018949599024000 \\1467009581041527168 & 1469066526780349312 & 1471944223587079808 & 1472043763749113984 & 1475499116478182400 \\1476416280975128832 & 1479656885339494656 & 1481262412833531520 & 1484142927140574976 & 1485907299704041088 \\1498477363310839168 & 1500489435229902976 & 1501129694594779776 & 1501277853786344704 & 1512783143459215744 \\1516361572771506176 & 1516500901510616960 & 1522392875086096384 & 1527475951701753984 & 1529110474518977920 \\1532216221205478272 & 1533286423976243584 & 1534269318651955072 & 1543128667952320384 & 1543778891645673216 \\1544941144151976832 & 1550080089702483584 & 1559682880662202240 & 1561486148450638080 & 1576029045153304448 \\1590912446864098688 & 1595706764237530624 & 1601706279498493184 & 1601734905456540160 & 1603695231610033024 \\1606083435288423680 & 1609793702916552448 & 1616751038836648192 & 1618407796700425856 & 1621470761217916800 \\1639946061258413312 & 1651212997426399872 & 1659006954318690688 & 1661132860050868096 & 1680352357664352384 \\1687099201530228352 & 1688495581297486336 & 1740372429681699712 & 1741837288407560192 & 1746823363186399104 \\1770226575557939840 & 1785179585801889792 & 1880722993023978240 & 1897394608662403584 & 1909569058536197760 \\1915730687339037056 & 1920542906137425664 & 1927742920592283264 & 1959250663239949568 & 1969372801649974656 \\2075971480449027840 & 2081319509311902336 & 2090990366909021312 & 2107177716389487744 & 2113756098755468288 \\2118428129820961152 & 2119154219811372928 & 2143783830030463744 & 2152492992912794112 & 2223355928911050752 \\2268048503896398720 & 2290361477475441664 & 2322233192826733184 & 2344587054493724288 & 2355425387285206016 \\2373208853993082112 & 2416023871138662784 & 2417033428971348608 & 2436947439975372672 & 2438115774159097856 \\2447968154259005952 & 2463622347979629952 & 2484686482506940416 & 2492679893385490944 & 2493751577920673664 \\2497639347957300096 & 2515939172813427840 & 2518385517465704960 & 2519551171589880192 & 2533982360488142336 \\2548084666562636544 & 2549249942728874368 & 2556488440091507584 & 2562681954730905344 & 2564879259999378560 \\2565036284003715712 & 2565305629992853376 & 2568434290329701888 & 2577125551790493824 & 2577340815551111680 \\2577877377225342464 & 2585460292309803264 & 2587112141027257472 & 2593008821887179008 & 2593099428517326848 \\2597454555419529728 & 2604228169817599104 & 2610361657994140032 & 2652715636170048384 & 2658069914200047872 \\2670534149811033088 & 2681582248805460864 & 2685833132557398656 & 2702017668840521088 & 2710316816966218496 \\2715380235515703168 & 2716342071966879744 & 2731449911488518144 & 2747904614098575488 & 2753048786625056640
\\\hline                                   %inserts single line
\end{tabular}
\end{table*}

\begin{table*}
\caption{{\it Gaia} DR2 source\_ids for selection B - extended }             % title of Table
\label{AppTable:1}      % is used to refer this table in the text
\centering                          % used for centering table
\begin{tabular}{c c c c c}       % centered columns (5 columns)
\hline\hline                 % inserts double horizontal lines
IDs members 1-105 & IDs members 106-210 & IDs members 211-315  & IDs members 316-419 & IDs members 420-523 \\    % table heading 
\hline % inserts single horizontal line
2753174783785251328 & 2762507266682585728 & 2763829601214919040 & 2777125789169639296 & 2777896134503915648 \\2782868607121209344 & 2807804328248244224 & 2816029465498234368 & 2838661293152878976 & 2841291982098062208 \\2865194471533729152 & 2865540508457717376 & 2866048379751327616 & 2869555134648788736 & 2876439211309388672 \\2882203332298969728 & 2890470899529272832 & 2891152566675457280 & 2901753198796969216 & 2902505745786910080 \\2954185571135157504 & 2959451922593403904 & 2964823797107665664 & 3074369755487517184 & 3085891537839264896 \\3085891537839267328 & 3124268582457078400 & 3140633331268375296 & 3165682645690499840 & 3179127443812906880 \\3191801342547132544 & 3202308378739431936 & 3214420461393486208 & 3233974932096638592 & 3236724913756362624 \\3244220864341745920 & 3246220914649655552 & 3258304272560352128 & 3267948604442696448 & 3270247545819617792 \\3275579971053642624 & 3283448591659397504 & 3302997083768142464 & 3306026508883214080 & 3311451812788556800 \\3318466112860232832 & 3344421699741253376 & 3360259919923274496 & 3368025671769397120 & 3368344461421835776 \\3375760083236221568 & 3435000842026369280 & 3458106907086594304 & 3500260464905928832 & 3515304597176877056 \\3520836313191203584 & 3612787302391075968 & 3638008690382625408 & 3638308577883569024 & 3642005964905520384 \\3645723619877700352 & 3649703023041037696 & 3665517234359034752 & 3667980999398592384 & 3685879227632832640 \\3693613948337587456 & 3694894948103300608 & 3696531536801618560 & 3697949460124343296 & 3697998658974002944 \\3698500933925098496 & 3700419135039744896 & 3703984129693916928 & 3705595086027166976 & 3710778767956154112 \\3712313170791794944 & 3712995761354141184 & 3716710186510880256 & 3718269255344140288 & 3719989475645411712 \\3726338021425401728 & 3734341989334114304 & 3736775002407500416 & 3738979076544369920 & 3742101345970116224 \\3742977210060826496 & 3767256797623849984 & 3773582902198012800 & 3784512979787195008 & 3786952418131994880 \\3798431869281771264 & 3799979225739262592 & 3803865655746018176 & 3805053368822108160 & 3808640525507460864 \\3812217030674464512 & 3812973391595125888 & 3815223026745362176 & 3817071305791426560 & 3830089282946013952 \\3838162172195153280 & 3842793349531132928 & 3845801334171290240 & 3868120909114479104 & 3875684174723979904 \\3887002066384290048 & 3892735332329204224 & 3892848582026608256 & 3894054127806485248 & 3897623005111511936 \\3902635335025097344 & 3916126823734155264 & 3921020612550678784 & 3922882360614215808 & 3925374438078436736 \\3927614040185794816 & 3928019553818870528 & 3928225540448909568 & 3929521177462206208 & 3932171034845348864 \\3933991448143753344 & 3935464175249715328 & 3937879149461334784 & 3939125274092506880 & 3950135439936325376 \\3959443664858300160 & 3960200884772486912 & 3961677219651037568 & 3965991806357554304 & 3966187038390853376 \\3971115633621943424 & 3986709182404800640 & 3988327319923520384 & 3997068162486541056 & 4001150546081363584 \\4008001534314802432 & 4018583985839228032 & 4024489531511197440 & 4027749961445908992 & 4317205919454821504 \\4349990744807366144 & 4418988157460172288 & 4422328435129459584 & 4423061744962215680 & 4440446153372208640 \\4457430485583460352 & 4466157103214240256 & 4494936029800093312 & 4534526935261529216 & 4561199025759521920 \\4564066449004092928 & 4573372234384968192 & 4607911055710222848 & 4669519677214249856 & 4670120873852223488 \\4679806226966950912 & 4692787168619498496 & 4731023544469431680 & 4738898315466921216 & 4754904073734461952 \\4768015406298936960 & 4790679540699217152 & 4913172141123403648 & 4919544021460799232 & 4928561700436553856 \\4963591625501132288 & 4965093901982556416 & 4987794109812160384 & 4988194744360498816 & 4998741805354135552 \\5017192164520193536 & 5032050552340352384 & 5048696058874427008 & 5049085217270417152 & 5067943490953707008 \\5079910055818560640 & 5122647149372975360 & 5155440732210734080 & 5167090027842913792 & 5187537405066478720 \\5195968563309851136 & 5204327871039930752 & 5280057494615723648 & 5283335207500744832 & 5284231275125726336 \\5296305287178801280 & 5340742599399608832 & 5360274938102023680 & 5366111940399568640 & 5388612346343578112 \\5392578250427878912 & 5414925927344979328 & 5446633059547018496 & 5532006972050819584 & 5539345181373346560 \\5552366427000089600 & 5558256888748624256 & 5563244926326975872 & 5621154039103070592 & 5658669238397675904 \\5723101272620828800 & 5753657422310407808 & 5806384261908917248 & 5808577237852977664 & 5810465477272078208 \\5818849597035446528 & 5840981323002935168 & 5851142528480811648 & 5897030680576289792 & 5907435045555029120 \\5916061474500671232 & 5926846098713325440 & 5927992614459539584 & 6050982889930146304 & 6094133670436009984 \\6109742818544552960 & 6144536275589632128 & 6170808423037019904 & 6221846137890957312 & 6255192092182279808 \\6258000485397994880 & 6322846447087671680 & 6330365869671479296 & 6330586119889886336 & 6336613092877645440 \\6356861870813950592 & 6387975296804951936 & 6393027243496440448 & 6485587710031536896 & 6490161850203676416 \\6492720894795089280 & 6499030270473444096 & 6545771884159036928 & 6558166197703420928 & 6583674450854788096 \\6615661065172699776 & 6620904533047264768 & 6626664771385047168 & 6695586523901966336 & 6701970872532935552 \\6713733658378414848 & 6778212250043037184 & 6780343958276205824 & 6793770167779622144 & 6857300327589197696 \\6866850891748542336 & 6909135394530431104 & 6914409197757803008\\\hline                                   %inserts single line
\end{tabular}
\end{table*}

\end{document}